%% file: SG-fullpaper.tex
\documentclass[aps,prb,twocolumn,showpacs,superscriptaddress]{revtex4}

\usepackage{graphicx}
\usepackage{amsmath,amssymb}

\usepackage[usenames, dvipsnames]{color}
\usepackage{amssymb}

\def\c{\ensuremath{\chi}}
\def\nc{\ensuremath{\chi_3}}

\def\ncb{\ensuremath{\overline{\chi}_{3}}}

\newcommand{\chiSG}{\ensuremath{\chi_\mathrm{SG}}}
\newcommand{\TSG}{\ensuremath{T_\mathrm{f}}}
\newcommand{\Tc}{\ensuremath{T_\mathrm{c}}}
\newcommand{\TN}{\ensuremath{T_\mathrm{N}}}
\newcommand{\Ns}{\ensuremath{N_\mathrm{s}}}
\newcommand{\qEA}{\ensuremath{q_\mathrm{EA}^2}}

\newcommand{\vQ}{\ensuremath{\vec{Q}}}

\newcommand{\Jcoop}{\ensuremath{J_2^\mathrm{coop}}}

\begin{document}
\title{Effect of magnetoelastic coupling on spin-glass behavior in Heisenberg pyrochlore antiferromagnets with bond disorder}
\author{Hiroshi Shinaoka} 
\affiliation{NRI, AIST, Tsukuba 305-8568, Japan}
\author{Yusuke Tomita} 
\affiliation{College of Engineering, Shibaura Institute of Technology, Minuma-ku, Saitama 330-8570, Japan}
\author{Yukitoshi Motome} 
\affiliation{Department of Applied Physics, University of Tokyo, 7-3-1 Hongo, Bunkyo-ku, Tokyo 113-8656, Japan}
\date{\today}

\begin{abstract}
Motivated by puzzling aspects of spin-glass behavior reported in frustrated magnetic materials,
we theoretically investigate effects of magnetoelastic coupling in geometrically frustrated classical spin models.
In particular, we consider bond-disordered Heisenberg antiferromagnets on a pyrochlore lattice coupled to local lattice distortions.
By integrating out the lattice degree of freedom, we derive an effective spin-only model, the bilinear-biquadratic model with bond disorder.
The effective model is analyzed by classical Monte Carlo simulations using an extended loop algorithm.
First, we discuss the phase diagrams in detail by showing the comprehensive Monte Carlo data for thermodynamic and magnetic properties. 
We show that the spin-glass transition temperature $\TSG$ is largely enhanced by the spin-lattice coupling $b$ in the weakly disordered regime.
By considering the limit of strong spin-lattice coupling, this enhancement is ascribed to the suppression of thermal fluctuations in semidiscrete degenerate manifold formed in the presence of the spin-lattice coupling.
We also find that, as increasing the strength of disorder $\Delta$, the system shows a concomitant transition of the nematic order and spin glass at a temperature determined by $b$, being almost independent of $\Delta$.
This is due to the fact that the spin-glass transition is triggered by the spin collinearity developed by the nematic order. 
Although further-neighbor exchange interactions originating in the cooperative lattice distortions result in the spin-lattice order in the weakly disordered regime, the concomitant transition remains robust with $\TSG$ almost independent of $\Delta$.
We find that the magnetic susceptibility shows hysteresis between the field-cooled and zero-field cooled data below $\TSG$,
and that the nonlinear susceptibility shows a negative divergence at the transition.
These features are common to the conventional spin-glass systems. 
Meanwhile, we find that the specific heat exhibits a broad peak at $\TSG$, and that the Curie-Weiss temperature varies with $\Delta$, even in the region where $\TSG$ is insensitive to $\Delta$. 
In addition, we clarified that the concomitant transition remains robust against a substantial external magnetic field.
These features are in clear contrast to the conventional spin-glass behavior.
Furthermore, we show that the cubic susceptibility obeys a Curie-Weiss-like law and the estimated ``Curie-Weiss'' temperature gives a good measure of the spin-lattice coupling even in the presence of bond randomness. 
We also show, by studying single-spin-flip dynamics in the nematic phase, that the spin freezing with rather high $\TSG$ may be practically observed in a realistic situation for weak disorder.
All these results are discussed in comparison with experiments for typical pyrochlore magnets, such as Y$_2$Mo$_2$O$_7$ and ZnCr$_2$O$_4$.
\end{abstract}


\maketitle

\input{introduction.tex}

\input{model.tex}

\input{MC.tex}

\input{phase_diagram.tex}
\input{magnetic_susceptibility.tex}

\input{magnetic_field.tex}

\input{spin_dynamics.tex}

\input{comparison.tex}
\input{summary.tex}
\appendix
\input{appendix.tex}

\end{document}

%% file: introduction.tex
\section{Introduction}
In magnets, competition between magnetic interactions suppresses formation of a simpleminded long-range order, and opens the possibility of unconventional magnetic behavior, such as an unexpected ordering, glassy behavior, and liquid-like state.
There are two major sources of such magnetic competition: randomness in the magnetic interactions and geometrical frustration of the lattice structures.~\cite{Ramirez94}

Randomness typically appears in the form of spatially random distribution of the strength and sign of magnetic interactions. 
Sufficiently strong randomness prevents the system from forming a long-range magnetic order, and instead, induces a new magnetic state called spin glass (SG).
A spin glass state is a disordered state in which spins are frozen randomly without any spatial periodicity.~\cite{Binder86}
It is distinguished from the paramagnetic state by the dynamical freezing of spin moments.
A typical example of SG is found in dilute magnetic metallic alloys,
in which randomly distributed magnetic moments interact with each other via the Ruderman-Kittel-Kasuya-Yosida (RKKY) interaction.~\cite{Ruderman54,Kasuya56,Yoshida57a,Yoshida57b}
The RKKY interaction is long-ranged and oscillating (changing the sign) with distance, and hence, the magnetic sector of the system can be mapped onto a localized spin model with random exchange couplings being both ferromagnetic and antiferromagnetic. 
Such randomness in the magnetic interactions is responsible for the SG behavior in these compounds.

On the other hand, geometrical frustration describes the competition arising from the geometry of lattice structures. 
It occurs even in the case in which the system is translationally invariant and magnetic interactions are not spatially random. 
A typical example is the Ising antiferromagnet on a triangular lattice.
In this model, it is impossible to satisfy all three nearest-neighbor antiferromagnetic interactions in every triangle. 
The frustration suppresses long-range ordering and leads to a disordered ground state with macroscopic degeneracy when the system has the nearest-neighbor interactions only.~\cite{Wannier50,Wannier73,Houtappel50,Husimi50}
Such a degenerate ground-state manifold is extremely sensitive to perturbations, such as small additional further-neighbor interactions and an external magnetic field. 
The macroscopic degeneracy and the sensitivity to perturbations are the source of unconventional magnetic behavior.~\cite{Diep05}

In the last two decades, the systems which include both two sources of magnetic competition, randomness and geometrical frustration, have been attracting growing interest.
Experimentally, SG behavior is widely seen in many magnets with geometrical frustration, ranging from quasi-two-dimensional systems such as SrCr$_8$Ga$_4$O$_{19}$~\cite{Ramirez90} to three-dimensional systems such as cubic spinels~\cite{Martinho01,Tristan05} and pyrochlores.~\cite{Munekata06,Zhou08,Gardner10}
While randomness inevitably existing in real materials might be relevant to the SG behavior, it has been intensively argued to what extent the geometrical frustration plays a role.
Specifically, on the theoretical side, it is still controversial if the geometrical frustration alone can induce SG behavior. 
Thus, it is desirable to study the effect of geometrical frustration by controlling the randomness.
It is also intriguing how the SG behavior in geometrically frustrated magnets is different from the canonical one driven solely by randomness.

To address these issues, we here focus on a typical geometrically frustrated system, 
an antiferromagnet on a pyrochlore lattice. 
As shown in Fig.~\ref{fig:model}(a), the pyrochlore lattice consists of a three-dimensional network of corner-sharing tetrahedra.
Antiferromagnets on the pyrochlore lattice are strongly frustrated.
For example, when considering classical Heisenberg spins with nearest-neighbor exchange interactions, 
no long-range ordering occurs down to zero temperature ($T$), and the ground state has continuous macroscopic degeneracy~\cite{Reimers92, Moessner98a} (see Sec.~\ref{subsec:no_disorder} for details).
Recently, the effect of randomness in the exchange interactions was studied on such extensively degenerate manifold.~\cite{Bellier-Castella01, Saunders07, Andreanov10}
It was found that the randomness immediately lifts the degeneracy and induces a SG transition.
The transition temperature $\TSG$ is proportional to the disorder strength $\Delta$ as $\TSG\propto \Delta$ in the weakly disordered regime. 
This implies that, in general, degenerate manifolds in geometrically frustrated magnets are sensitive to randomness, potentially possessing an instability toward SG. 
This gives a clue to explain why SG is prevailing in geometrically frustrated materials.

However, several characteristics of the SG in geometrically frustrated magnets still remain puzzling.
Insulating molybdate pyrochlores $R_2$Mo$_2$O$_7$ ($R=\mathrm{Dy,~Tb,~Gd,~Lu}$) are typical SG materials with geometrical frustration~\cite{Greedan86,Gingras96,Gingras97,Silverstein13,Clark14}.
In these compounds, the magnetic Mo$^{4+}$ cations constitute a pyrochlore lattice.
Among them, Y$_2$Mo$_2$O$_7$ is one of the most intensively-studied compounds for its SG behavior.
The compound exhibits a SG transition at $\TSG\simeq 22 \text{K}$ which is identified by a bifurcation of field-cool (FC) and zero-field-cool (ZFC) magnetic susceptibilities.~\cite{Greedan86}
The SG behavior resembles that of the canonical SG theory at first glance:
the transition is second order and the nonlinear susceptibility $\chi_3$ shows a negative divergence.~\cite{Gingras97}
Furthermore, the estimated critical exponents do not contradict with those of the canonical ones.~\cite{Gingras97}
There are, however, several aspects that cannot be explained by the conventional SG theory.
One concerns the critical temperature $\TSG$. 
$\TSG$ remains unchanged for the substitution of Y$^{3+}$ by La$^{3+}$ up to 50\%, 
despite random lattice distortions induced by the substitution and a substantial increase of the Curie-Weiss temperature $\theta_\mathrm{CW}$.~\cite{Sato87}
This indicates that $\TSG$ does not strongly depend on either the randomness $\Delta$ or the dominant magnetic interactions. 
Moreover, $\TSG$ appears to be much higher than 
that theoretically expected for a moderate strength of disorder $\Delta$; 
e.g., the experimental value is about 20--30 times higher than a numerical estimate of $\TSG/J\simeq 0.01$ for a nearest-neighbor Heisenberg antiferromagnet with $\Delta/J=0.1$.~\cite{Saunders07, Andreanov10,Tam10}
Another unconventional aspect is the specific heat. 
In Y$_2$Mo$_2$O$_7$, a broad peak is observed in the specific heat at $\TSG$.~\cite{Raju92,Silverstein13} 
This is in contrast to the canonical SG which has no clear anomaly at $\TSG$, except for a broad hump at a higher temperature.~\cite{Binder86}
The last but not least is the robustness against an external magnetic field. 
The peak in the specific heat as well as the bifurcation of the FC and ZFC susceptibilities is almost unaffected by a magnetic field up to several Tesla.~\cite{Silverstein13}
This is also in contrast to the canonical SG which is strongly disturbed by the magnetic field.~\cite{Binder86,Young04,Sasaki07}

Similar puzzling SG behavior, in particular, the insensitive $\TSG$, is observed in other frustrated magnets, e.g., spinel oxides (Zn$_{1-x}$Cd$_x$)Cr$_2$O$_4$.
In this case also, the magnetic Cr$^{3+}$ cations comprise a pyrochlore lattice.
The stoichiometric compound with $x=0$ exhibits a long-range antiferromagnetic order accompanied by a lattice distortion at $T_\mathrm{c}\simeq $ 13 K.~\cite{Kino71,Ratcliff02}
The order, however, is destroyed by a small amount of Cd substitution at $x\simeq 0.03$, and for larger $x$, the compounds exhibit SG behavior.~\cite{Ratcliff02}
In the SG region, the SG transition temperature $\TSG$ is weakly dependent on $x$; $\TSG$ remains $\simeq 10$ K up to $x\sim0.1$.
This also indicates the robustness of $\TSG$ against the randomness $\Delta$, as in (Y$_{1-x}$La$_x$)$_2$Mo$_2$O$_7$. 
Similar robust behavior of $\TSG$ is also seen in another spinel CoAl$_{2}$O$_{4}$, in which magnetic Co$^{3+}$ cations form a diamond lattice,
while changing the fraction of intersite mixing between Co and nonmagnetic Al sites.~\cite{Hanashima13}
In this case, although the diamond lattice is bipartite, the frustration may come from the competition between the nearest- and second-neighbor interactions.

These experimental results indicate that the SG transition temperature $\TSG$ is not set by the strength of randomness $\Delta$, but by another energy scale.
In other words, some important factor is missing in the previous theories, in which $\TSG$ was predicted to be proportional to $\Delta$.~\cite{Bellier-Castella01, Saunders07, Andreanov10,Tam10}
A possible candidate for the missing energy scale is the magnetoelastic coupling to local lattice distortions.
Indeed, the importance of local lattice distortions has been pointed out for Y$_2$Mo$_2$O$_7$ by various microscopic probes such as x-ray-absorption fine-structure (XAFS) technique,~\cite{Booth00} neutron pair distribution function analysis,~\cite{Greedan09} nuclear magnetic resonance (NMR),~\cite{Keren01,Ofer10} and muon spin rotation and relaxation ($\mu$SR) techniques.~\cite{Sagi05,Ofer10}
Meanwhile, the importance of the magnetoelastic coupling in (Zn$_{1-x}$Cd$_x$)Cr$_2$O$_4$ is obvious as the compound at $x=0$ shows the spin-lattice coupled ordering.~\cite{Kino71}
Theoretically, it was shown that the randomness in the strength of magnetic interactions destroys the spin-lattice order and induces a SG state.~\cite{Saunders08}
However, the argument was limited to a uniform global lattice distortion, and $\TSG$ was deduced to behave similarly to the case in the absence of the magnetoelastic coupling, i.e., $\TSG\propto \Delta$, after the uniform lattice distortion is destroyed.

Motivated by the puzzling SG behavior and the implication of magnetoelastic coupling, the authors recently investigated the SG behavior in bond-disordered classical Heisenberg antiferromagnets on the pyrochlore lattice.~\cite{Shinaoka10b,Shinaoka-LT2011}
The main conclusion was that the spin-lattice coupling enhances $\TSG$, and induces a concomitant transition with nematic order and spin glass. 
In this concomitant transition, $\TSG~(=\Tc)$ becomes almost independent of $\Delta$.
The results give a reasonable account of the puzzling behavior of $\TSG$ in the frustrated magnets.

The aim of the present paper is to provide a comprehensive description of the characteristic properties of the SG transition in 
pyrochlore antiferromagnets coupled with local lattice distortions.
For the comparison with experiments in a broader viewpoint than in the previous studies,~\cite{Shinaoka10b,Shinaoka-LT2011} we investigate  
thermodynamic and magnetic observables, such as the specific heat, spin collinearlity, SG susceptibility, and sublattice magnetization,
by systematically controlling the bond randomness, temperature, and magnetic field.
We show the detailed analyses of the phase diagrams and critical properties; the tables for the critical temperatures and exponents are presented.
For the linear magnetic susceptibility, we show that the Curie-Weiss temperature is dependent on the strength of bond randomness as well as the temperature range for the fitting. 
We also find that it exhibits a bifurcation between the FC and ZFC measurements below the concomitant transition temperature. 
From the analysis of the nonlinear magnetic susceptibility,
we find that it shows a negative divergence at the concomitant transition, whereas it is positively divergent at the nematic transition. 
In addition, we show that the cubic susceptibility obeys a Curie-Weiss-like law and the estimated Curie-Weiss temperature gives a good measure of the spin-lattice coupling even in the presence of bond randomness.
The hysteresis in the magnetic susceptibility and the negative divergence of the nonlinear susceptibility are consistent with the experimental results in Y$_2$Mo$_2$O$_4$.~\cite{Greedan86,Gingras97}
We also clarify effects of an external magnetic field on the specific heat and magnetic susceptibility.
We find that, in sharp contrast to the conventional SG, the transition is robust against the magnetic field.
This is also consistent with the experimental results.~\cite{Silverstein13}
Finally, we show that spin relaxation suffers from severe dynamical freezing in the nematic phase due to the spin-ice type manifold even when $\Tc > \TSG$.
This suggests that SG with rather high $\TSG$ may occur in a realistic situation even for an extremely weak disorder.

This paper is organized as follows.
In Sec.~II, we introduce the models studied in this paper with qualitative arguments on the phase diagrams. 
In Sec.~III, we describe the classical MC method used for the present study.
In Sec.~IV, we show the results on the phase diagrams obtained by MC simulation.
In Sec.~V, we investigate linear and nonlinear magnetic susceptibilities.
In Sec.~VI, we discuss effects of an external magnetic field on the specific heat and the magnetic susceptibility.
In Sec.~VII, we investigate single-spin-flip spin relaxation in the nematic phase. 
In Sec.~VIII, we discuss our theoretical results in comparison with existing experimental results.
Summary is given in Sec.~IX.

%% file: model.tex
\section{Model}
In this section, we introduce the microscopic models studied in this paper.
In Sec.~\ref{sec:model1}, we introduce an antiferromagnet on a pyrochlore lattice coupled to local lattice distortions.
In Sec.~\ref{sec:model2}, we show the derivation of effective spin-only models by integrating out the lattice degree of freedom, whose procedure was described only briefly in our previous paper.~\cite{Shinaoka10b}
In Sec.~\ref{sec:model3}, we present qualitative arguments expected for the phase diagrams of the effective spin-only models,
and show what we will clarify in the rest of the present paper.

\subsection{Pyrochlore antiferromagnet coupled to local lattice distortions}\label{sec:model1}
To consider effects of spin-lattice coupling,
we start with a classical Heisenberg antiferromagnet coupled with lattice distortions; 
\begin{equation}
	\mathcal{H}= \sum_{\langle i,j \rangle} 
	\Big[ J_{ij} \left(1-\alpha \rho_{ij}\right)\vec{S}_i \cdot \vec{S}_j + \frac{K}{2}\rho_{ij}^2 
	\Big],\label{eq:bigHam}
\end{equation}
where $\vec{S}_i$ ($|\vec{S}_i|=1$) denotes a Heisenberg spin at site $i$,
and the sum runs over nearest-neighbor bonds of the pyrochlore lattice [Fig.~\ref{fig:model}(a)].
Here, $\rho_{ij}$ is the change in distance between nearest-neighboring spins $\vec{S}_i$ and $\vec{S}_j$, relative to the equilibrium lattice constant; we treat the distortions as classical objects and neglect the kinetic energy of phonons.
The model incorporates the magnetoelastic coupling up to the linear order of bond distortion $\rho_{ij}$.
We take the coupling constant $\alpha$ being positive; namely, the exchange interaction is enhanced on a shorter bond than a longer bond, as illustrated in Fig.~\ref{fig:model}(b).
In addition to the magnetoelastic coupling, we introduce quenched randomness in the coupling constant $J_{ij}$ as an extrinsic bond disorder. 
Here, we assume the distribution of $J_{ij}$ to be uniform as
\begin{eqnarray}
	J_{ij} &\in& [J-\Delta,J+\Delta]\label{eq:Jij}
\end{eqnarray}
with $0\le \Delta < J$.
Consequently, all the exchange couplings are antiferromagnetic in the model (\ref{eq:bigHam}), 
while the amplitudes are modulated by both magnetoelastic coupling and quenched disorder.
The last term in Eq.~(\ref{eq:bigHam}) represents the elastic energy of lattice distortions in the harmonic approximation ($K>0$).
Hereafter, Boltzmann constant $k_\mathrm{B}$ is set to unity, and all the energy scales including $T$ are measured in units of $J$.

\subsection{Effective spin model by integrating out lattice degrees of freedom}\label{sec:model2}
In general, the lattice distortions $\rho_{ij}$ depend on each other through, e.g., the movement of an ion shared by two neighboring bonds and a long-range strain effect.
Such cooperative aspect may lead to spin-lattice ordering in which 
a structural transition and magnetic ordering take place in a coupled manner.  
The spin-lattice ordering in ZnCr$_2$O$_4$ is a typical example.~\cite{Kino71,Ratcliff02}
When the cooperative aspect is less important and can be ignored, 
the model (\ref{eq:bigHam}) is much simplified; integrating out $\rho_{ij}$ by completing the squares, we end up with the spin-only model,
\begin{equation}
	\mathcal{H}= \sum_{\langle i,j \rangle} 
	\Big[ 
	J_{ij} \vec{S}_i \cdot \vec{S}_j - b_{ij} \big( \vec{S}_i \cdot \vec{S}_j \big)^2 
	\Big]. \label{eq:Ham}
\end{equation}
The second term describes the biquadratic coupling generated by the coupling to local lattice distortions.
It tends to align the direction of spins (but not the orientation), i.e., favors spin collinearity. 
Here, $b_{ij}~(\equiv J_{ij}^2 \alpha^2/2K >0)$ is the biquadratic coupling constant, which is also a bond-disordered variable.
Hereafter we use $b\equiv\alpha^2/2K$ as a parameter which measures the strength of the spin-lattice coupling.
The model (\ref{eq:Ham}) is considered as a fundamental model to unveil intrinsic effects of the coupling to independent local lattice distortions.
We discuss the results in comparison with the experimental data for Mo pyrochlores which show no uniform lattice distortion.
\begin{figure}[h]
 \centering
 \includegraphics[width=.475\textwidth,clip]{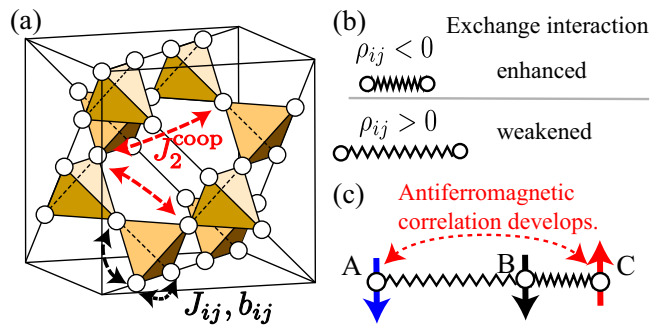}
 \caption{(Color online) 
 (a) 16-site cubic unit cell of the pyrochlore lattice.
 (b) Schematic illustration of the coupling of spins to a bond distortion $\rho_{ij}$.
 The antiferromagnetic exchange interactions are enhanced on shorter bonds.
 (c) Schematic illustration of a cooperative aspect of bond distortions; A shift of the B site while elongating (shortening) the AB (BC) bond enhances an antiferromagnetic spin correlation between the next nearest-neighbor spins A and C.
 }
 \label{fig:model}
\end{figure}

On the other hand, when the cooperative aspect of lattice distortions becomes important as in ZnCr$_2$O$_4$, it is necessary to include additional contributions beyond the model in Eq.~(\ref{eq:Ham}). 
Effects of the cooperative aspect were discussed in previous theoretical studies.~\cite{Tchernyshyov02,Tchernyshyov02b,Bergman06}
In particular, Bergman \textit{et al.} showed that a cooperative lattice distortion induces effective multiple-spin interactions.~\cite{Bergman06}
They also found that the multiple-spin interactions bring about effective further-neighbor interactions for collinear spin states that are favored by $b$ at low $T$.
In general, such effective exchange interactions are complicated and dependent on the details of materials.
Tchernyshyov \textit{et al.} showed that several N\'{e}el ordered phases, including collinear, coplanar, and noncoplanar ones, can appear 
as a result of cooperative couplings.~\cite{Tchernyshyov02,Tchernyshyov02b}
Indeed, Cr spinels $A$Cr$_2$O$_4$ exhibit a variety of different spin-lattice orderings for different cations $A$; e.g., Cr spinels $A$Cr$_2$O$_4$ 
show complex different $\vec{q}\neq 0$ coplanar magnetic orderings for $A=~$Zn~\cite{Ji09} and Hg,~\cite{Ueda06,Matsuda07} and noncollinear ordering for $A$=Cd.~\cite{Matsuda20077}
However, the study of material-dependent magnetic structures is out of the scope of the present study.
Our aim is to extract an intrinsic effect of the cooperative aspect. 
For the purpose, we take into account one of the simplest contributions, the effective antiferromagnetic interaction for second neighbors, $J_2^\mathrm{coop}$;
\begin{eqnarray}
	\mathcal{H}_\mathrm{coop} &\equiv& J_2^\mathrm{coop} \sum_{\langle\langle i,j \rangle\rangle} \vec{S}_i \cdot \vec{S}_j,\label{eq:J2}
\end{eqnarray}
where the sum is over the second neighbor pairs [see Fig.~\ref{fig:model}(a)].

The physical meaning of $J_2^\mathrm{coop}$ can be understood intuitively by considering two neighboring bonds, as shown in Fig.~\ref{fig:model}(c).
Once the center site is shifted toward one of the neighboring sites, antiferromagnetic spin correlations are enhanced on the shorter bond by the magnetoelastic coupling, while they are reduced on the other elongated one.
These two effects cooperatively enhance antiferromagnetic correlations between the second-neighbor spins, which is effectively represented by $J_2^{\rm coop}$.

\subsection{Qualitative arguments on the effective model}\label{sec:model3}
In the present study, we investigate effects of the spin-lattice coupling by using the bilinear-biquadratic model that incorporates Eq.~(\ref{eq:J2}) into Eq.~(\ref{eq:Ham}); 
\begin{eqnarray}
	\mathcal{H}&=&\sum_{\langle i,j \rangle} 
	\Big[ 
	J_{ij} \vec{S}_i \cdot \vec{S}_j - b \left(J_{ij}\right)^2 \big( \vec{S}_i \cdot \vec{S}_j \big)^2\Big]\nonumber \\ 
	  &&+J_2^\mathrm{coop}\sum_{\langle\langle i,j \rangle\rangle} \vec{S}_i \cdot \vec{S}_j. \label{eq:Ham2}
\end{eqnarray}
In this section, giving qualitative arguments on the expected phase diagram of the model (\ref{eq:Ham2}), we present our motivations in the current study.
\begin{figure}[h]
 \centering
 \includegraphics[width=.425\textwidth,clip]{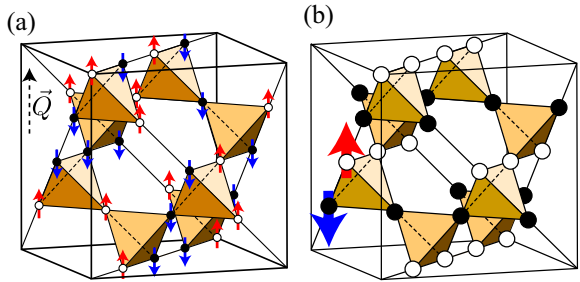}
 \caption{(Color online) 
 Ground states of the model (\ref{eq:Ham2}) for (a) $J_2^\mathrm{coop}=0$ and (b) $J_2^\mathrm{coop}>0$ ($b>0$ and $\Delta=0$).
 In (a), we show one of the macroscopically degenerate ground states (ice-rules configurations).
 The common axis of spins, which is denoted by a broken arrow $\vQ$, is spontaneously selected below $\Tc$.
 In every tetrahedron, two of spins are aligned parallel to $\vQ$ and the other two antiparallel to $\vQ$.
 In (b), we present the $\vec{q}=0$ spin-lattice (N\'{e}el) order.
 The open and filled circles denote two nonequivalent sites with opposite spins.
 }
 \label{fig:ordering-pattern}
\end{figure}

\subsubsection{In the absence of bond disorder}\label{subsec:no_disorder}
First, let us discuss the case in the absence of bond disorder, $\Delta=0$. 
When both $b$ and $J_2^\mathrm{coop}$ are zero, i.e., in the absence of spin-lattice coupling, 
the model is reduced to a simple antiferromagnetic Heisenberg model on the pyrochlore lattice with nearest-neighbor exchange interactions only.
The Hamiltonian is rewritten into
\begin{eqnarray}
\label{eq:H_wo_b}
	\mathcal{H}&=&\sum_{\langle i,j \rangle} \vec{S}_i \cdot \vec{S}_j =\frac{1}{2} \sum_t |\vec{M}_t|^2 + \mathrm{const.},
\end{eqnarray}
where $\vec{M}_t$ is the sum of four spin moments $\vec{S}_i$ on a tetrahedron $t$.
Thus, the ground state is identified by a collection of local constraints that $\vec{M}_t$ vanishes on every tetrahedron.
This set of constraints, however, does not 
select a unique ground state, leaving the continuous macroscopic degeneracy at $T=0$.~\cite{Reimers92,Moessner98a,Moessner98b}
In addition, thermal fluctuations do not induce any order.
Therefore, the system does not exhibits any magnetic ordering in the entire range of $T$.~\cite{Reimers92,Moessner98a,Moessner98b} 

\begin{figure}[ht]
 \centering
 \resizebox{0.4\textwidth}{!}{\includegraphics{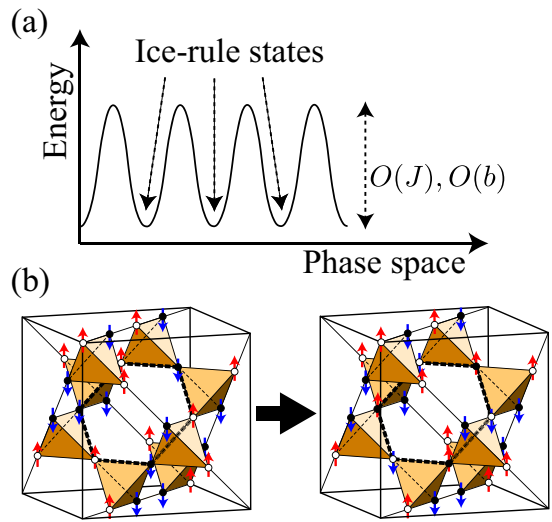}}
 \caption{(Color online) 
 (a) Schematic picture of a multiple valley structure in the spin-ice type manifold.
 (b) Two different ice-rule states are shown.
 The hexagon with a bold dashed line denotes one of the shortest loops on which a flip of all spins transforms the ice-rule state to another ice-rule state. See the text for details.
 }
 \label{fig:multivalley}
\end{figure}

For $b>0$ and $J_2^\mathrm{coop}=0$, the model exhibits a weak first-order transition at $\Tc \sim b$ to a nematic state.~\cite{Shannon10}
Below $\Tc$, all spins are aligned parallel or antiparallel to a spontaneously-selected axis $\vQ$.
This transition is not a magnetic ordering but a directional ordering of magnetic moments, corresponding to the ordering of spin quadrupole moments.
The ground state is now identified by a collection of local constraints equivalent to the so-called ice rule~\cite{Bernal33,Pauling35}; 
in every tetrahedron, two out of four spins are aligned parallel to each other and the other two are antiparallel to them --- `two-up two-down' (ice-rule) configuration as exemplified in Fig.~\ref{fig:ordering-pattern}(a).
The system still remains magnetically disordered down to $T=0$, while the ground-state degenerate manifold is modified to a \textit{semidiscrete} form due to the spin-lattice coupling $b$.
That is, the energy landscape has a multivalley structure 
in which the valleys correspond to different ice-rule configurations [see Fig.~\ref{fig:multivalley}(a)].
\begin{figure}[h]
 \centering
 \includegraphics[width=.49\textwidth,clip]{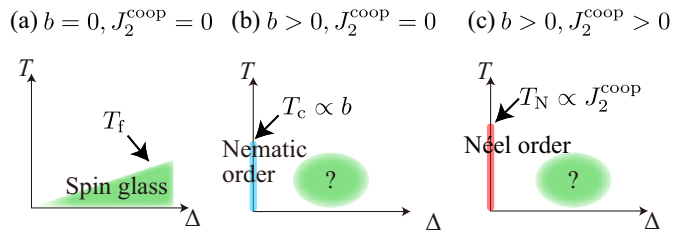}
 \caption{(Color online) 
 Schematic phase diagrams of model (\ref{eq:Ham2}) for three typical cases: (a) in the absence of the spin-lattice coupling $b=0$, (b) $b>0$ and $J_2^\mathrm{coop}=0$, and (c) $b>0$ and $J_2^\mathrm{coop}>0$.
Our interest here is how the SG transition is induced by the bond disorder $\Delta$ in the competition with the nematic and N\'{e}el orderings, as shown in (b) and (c). 
 }
 \label{fig:model-pd-schematic}
\end{figure}

When $J_2^\mathrm{coop}$ is turned on, the system exhibits a magnetic transition to a spin-lattice (N\'{e}el) ordered state at $\TN\propto J_2^\mathrm{coop}$ as a consequence of the lifting of the degeneracy.
The ordering pattern depends on the sign of $J_2^\mathrm{coop}$ in the absence of $b$.~\cite{Reimers91,Chern08}
The antiferromagnetic $J_2^\mathrm{coop}>0$ induces the $\vec{q}=0$ collinear four-sublattice N\'{e}el order illustrated in Fig.~\ref{fig:ordering-pattern}(b),
while the ferromagnetic $J_2^\mathrm{coop}<0$ induces a multiple-$q$ order.~\cite{Chern08}

In the following, we focus on the case with $J_2^\mathrm{coop}>0$.
When $b>0$, the antiferromagnetic $J_2^\mathrm{coop}$ also selects the $\vec{q}=0$ collinear order from the semidiscrete manifold.
The system will exhibit two successive transitions in the small $J_2^\mathrm{coop}$ region, 
the nematic transition at $T_\mathrm{c}$ and the N\'{e}el transition at lower $\TN$.
When $J_2^\mathrm{coop}$ becomes sufficiently large, the nematic phase will be completely taken over by the spin-lattice order, 
and hence, the system will exhibit only a single transition at $\TN$. 
A similar situation was studied for a third-neighbor ferromagnetic interaction.~\cite{Shannon10}

\subsubsection{Effects of bond disorder: Motivation of the present study}\label{sec:motivation}
We are interested in how SG appears when the bond disorder $\Delta$ is turned on.
In the absence of the spin-lattice coupling,
the bond disorder $\Delta$ induces effective long-range interactions and lifts the ground-state degeneracy.~\cite{Saunders07} 
Consequently, a SG transition at a finite $T$ is induced immediately by switching on $\Delta$; 
the transition temperature $\TSG$ is proportional to the strength of disorder $\Delta$ in the small $\Delta$ region.~\cite{Saunders07, Andreanov10}
This is schematically shown in Fig.~\ref{fig:model-pd-schematic}(a).
The value of $\TSG$ was estimated as $\TSG=0.02$--0.032 at $\Delta=0.1$ in the previous MC studies.~\cite{Saunders07, Andreanov10}

For $b>0$ and $J_2^\mathrm{coop}=0$, a SG may appear immediately for $\Delta>0$ because the ground states are also macroscopically degenerate.
However, the energy landscape has a multiple valley structure and the ground-state manifold is now semidiscrete.
Furthermore, the system has a new energy scale set by the spin-lattice coupling $b$.
Therefore, it is highly nontrivial how $\TSG$ appears and develops as a function of $\Delta$ [see Fig.~\ref{fig:model-pd-schematic}(b)].

On the other hand, for $J_2^\mathrm{coop}>0$, the spin-lattice order is induced by the cooperative coupling at $\Delta=0$.
In this case, a SG appears in competition with the spin-lattice order, as observed in ZnCr$_2$O$_4$. 
Although the effect of a uniform lattice distortion was studied in the previous theoretical work,~\cite{Saunders08} 
it is unclear what happens in the case with local lattice distortions [see Fig.~\ref{fig:model-pd-schematic}(c)].

Our motivation is, therefore, to clarify the SG behavior induced by the quenched bond disorder $\Delta$ in the presence of the spin-lattice coupling $b$. 
We clarify the $\Delta$-$T$ phase diagrams by extensive MC simulations for the two cases: (i) $b>0$ and $J_2^\mathrm{coop}=0$ [Fig.~\ref{fig:model-pd-schematic}(b)] and (ii) $b>0$ and $J_2^\mathrm{coop}>0$ [Fig.~\ref{fig:model-pd-schematic}(c)]. 
The corresponding MC results are shown in Figs.~\ref{fig:pd}(a) and \ref{fig:pd}(b).
We will discuss the results for the former case in comparison with the experiments in $R_2$Mo$_2$O$_7$ in which no structural transition is observed in even in high-quality stoichiometric samples. 
Meanwhile, we compare the latter with (Zn$_{1-x}$Cd$_x$)Cr$_2$O$_4$ in which the spin-lattice order at $x=0$ is destabilized and taken over by SG.

%% file: MC.tex
\section{Monte Carlo method}\label{sec:MC}
In the following sections, we investigate thermodynamic properties of the model (\ref{eq:Ham2}) using classical MC simulation.
We use the conventional single-spin update~\cite{Marsaglia72} together with the overrelaxation update.~\cite{Alonso96}
We also adopt the exchange MC method~\cite{Hukushima96} for efficient sampling.
The single-spin flip dynamics, however, is severely suppressed by dynamical freezing at low $T$ below the nematic transition temperature $\Tc$ because of the spin-ice type local constraint (see Sec.~\ref{sec:spin-relax}).
Therefore, in order to ensure the ergodicity at low $T$, we also adopt a nonlocal update method called the loop algorithm.~\cite{Melko01,Melko04,Shinaoka-LM1,Shinaoka-LM2,Wang12}
After a brief review on the loop algorithm originally developed for Ising modes~\cite{Melko01,Melko04} in Sec.~\ref{sec:lm-ising},
we introduce an extended loop algorithm for Heisenberg spin systems which was recently developed by the authors~\cite{Shinaoka-LM1,Shinaoka-LM2} in Sec.~\ref{sec:lm-b}.
Section~\ref{sec:lm-detail} summarizes the flowchart of MC simulation with the extended loop algorithm.
We describe the system setup and the definitions of observables for MC simulation in Sec.~\ref{sec:model4}.

\subsection{Loop algorithm for Ising models with spin-ice type degeneracy}\label{sec:lm-ising}
For Ising models showing spin-ice type degeneracy in the ground state,
it is hard to clarify low-$T$ properties by single-spin-flip MC calculations.
This is because the single-spin-flip MC dynamics is frozen out at low $T\ll J$ due to ``multiple valley'' energy structure of degenerate ground-state manifold [see Fig.~\ref{fig:multivalley}(a)]; the low-energy ice-rule states are separated by large energy barriers, which are not able to overpass by any single-spin flip as it inevitably violates the ice rule.
This is in clear contrast to the nearest-neighbor Heisenberg antiferromagnet in Eq.~(\ref{eq:H_wo_b}), in which the ground-state manifold is continuously connected without any energy barrier; the degenerate states can be sampled over by single-spin flips down to low $T\ll J$.

The difficulty can be avoided by a nonlocal flip based on the loop algorithm.~\cite{Melko01,Melko04}
Let us consider an Ising model on the pyrochlore lattice with nearest-neighbor antiferromagnetic interactions [see Fig.~\ref{fig:multivalley}(b)].
This model has the macroscopically degenerate ground states that satisfy the two--up two--down constraint on every tetrahedron.~\cite{Anderson56}
The nonlocal flip, called the loop flip, consists of two steps:
first, we identify a closed loop which consists of alternating alignment of up and down spins, and next,
we flip all Ising spins on the loop. 
Such a loop update transforms an ice-rule state to another ice-rule state bypassing the energy barriers, as it does not cost the exchange energy.
Indeed, the loop algorithm has been successfully applied to the study of
low-$T$ properties of spin-ice type Ising models.~\cite{Melko01,Isakov04,Jacob05, Jaubert10}

\subsection{Extension of the loop algorithm to bilinear-biquadratic Heisenberg spin models}\label{sec:lm-b}
We have a similar difficulty for the bilinear-biquadratic model (\ref{eq:Ham2})
because the energy landscape also has a ``multiple valley'' structure below the nematic transition temperature $\Tc$.
The problem becomes serious as we need to determine the SG transition temperature $\TSG$ which is much lower than $\Tc$ in the small $\Delta$ region [see Fig.~\ref{fig:pd}(a)].

Recently, the authors extended the loop algorithm to Heisenberg spin systems with spin-ice type degeneracy:
Heisenberg models with single-ion anisotropy\cite{Shinaoka-LM1} and bilinear-biquadratic models.~\cite{Shinaoka-LM2}
We employ the latter in the following simulations.
In the extended algorithm, at each MC step, all spins are projected onto an axis to define a set of Ising discrete variables, and 
spins on a closed loop are flipped simultaneously, similar to the Ising case. 
In Ref.~\onlinecite{Shinaoka-LM2}, the authors tested the efficiency of three different ways of the loop flip. 
The acceptance rates of the three updates are affected by thermal fluctuations in different ways,
and therefore, the most efficient method depends on the value of $b$.
In the present study, we adopt \textit{rotate}, which is the loop update with a cyclic rotation of spins along the loop, as it has the highest acceptance rate at low $T$ for the value of $b=0.2$ used throughout the following simulations.

\subsection{Simulation details}\label{sec:lm-detail}
\begin{figure}[ht]
 \centering
 \resizebox{0.3\textwidth}{!}{\includegraphics{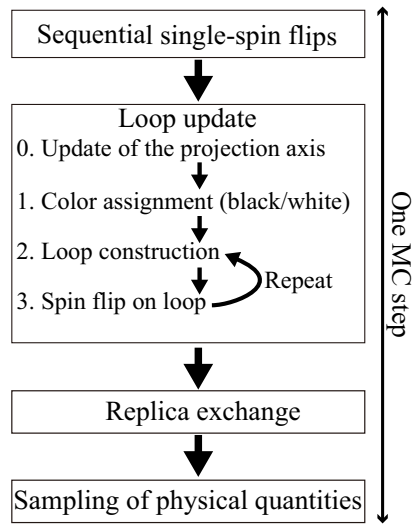}}
 \caption{(Color online)
 Flowchart of the MC simulation in this study.
 Each MC step consists of a lattice sweep by single-spin updates, loop flips, and a replica exchange between neighboring temperatures.
 }
 \label{fig:flowchart}
\end{figure}

We here describe technical aspects of MC simulations with the loop update.
As illustrated in Fig.~\ref{fig:flowchart},
each MC step consists of a sweep of the lattice by sequential single-spin flips, 
followed by the loop update and replica exchange between neighboring temperatures.

In the single-spin-flip sweep, on each site, 
we first try to update $\vec{S}_i$ to a randomly chosen new spin state.~\cite{Marsaglia72}
Then, we try to rotate the spin around the molecular magnetic field by an angle of $\pi$ (overrelaxation update).~\cite{Michael87,Alonso96}
These two updates are performed by the standard Metropolis algorithm sequentially.

In the extended loop algorithm, the projection spin axis is updated at every MC step being parallel to the common axis of spins $\vec{Q}$ in the nematic phase [see Fig.~\ref{fig:ordering-pattern}(a)].~\cite{Shinaoka-LM2}
We repeat the loop flip so that the total cpu time spent for the loop flips is comparable to that for
the single-spin-flip sweep.

For the replica exchange MC method, we optimize the distribution of temperature points in thermalization MC steps for each configuration of $\{J_{ij}\}$ so that the exchange rate is independent of $T$.
Thermodynamic observables are measured using the reweighting method.~\cite{LandauMC}

\subsection{System setup and observables}\label{sec:model4}
In the following MC simulations,
we consider finite size systems composed of $L^3$ cubic unit cells,
in which the total number of spins are $\Ns = 16L^3$, under periodic boundary conditions [see Fig.~\ref{fig:model}(a)].
We take the spin-lattice coupling $b=0.2$ throughout the present study.
The cooperative coupling is taken to be $J_2^\mathrm{coop}=0$ or $0.075$.

To identify the SG, nematic, and $\vec{q}=0$ collinear antiferromagnetic transitions, 
we calculate the SG susceptibility $\chiSG$, nematic order parameter $Q^2$, sublattice magnetization $m_{\text{s}}$, and specific heat $C$.
The SG susceptibility $\chiSG$ is given by
\begin{eqnarray}
\chiSG &\equiv& \Ns \qEA, 
\end{eqnarray}
where $\qEA$ is the Edwards-Anderson order parameter~\cite{Edwards75} for SG defined by 
\begin{eqnarray}
	\qEA &\equiv& \frac{1}{\Ns^2} \left\langle \left\langle \sum_{\mu,\nu=x,y,z} \left(\sum_{i=1}^{N_\mathrm{s}} S_{i\mu}^\alpha S_{i\nu}^\beta\right)^2 \right\rangle_T \right\rangle_\Delta.
	\label{eq:qEA}
\end{eqnarray}
Here $\langle \cdots \rangle_T$ denotes a thermal average and 
$\langle \cdots \rangle_\Delta$ a random average over the interaction sets $\{J_{ij}\}$; 
the upper suffixes $\alpha$ and $\beta$ denote two independent replicas of the system with the same interaction set.
$S_{i\mu}$ ($\mu=x,y,z$) are $x$,~$y$,~$z$-components of the normalized Heisenberg spin $\vec{S}_i$ at site $i$.

The nematic order parameter $Q^2$, which measures the spin collinearity, is defined as
\begin{eqnarray}
  Q^2 \equiv \frac{2}{\Ns^2} \left\langle \left\langle \sum_{i,j=1}^{\Ns} \left\{\left(\vec{S}_i \cdot \vec{S}_j\right)^2 - 
	\frac13 \right\} \right\rangle_T \right\rangle_\Delta.
\end{eqnarray}
Note that this is given by the summation of the quadrupole moments and invariant under $O(3)$ rotations.~\cite{Shannon10}
The susceptibility of $Q$, $\chi_Q$, is defined as
\begin{eqnarray}
	\chi_Q &\equiv& \Ns Q^2.
\end{eqnarray}

The linear magnetic susceptibility $\chi$ and the nonlinear magnetic susceptibility $\nc$ are defined by
\begin{eqnarray}
	\chi   &=& \frac{\partial m}{\partial H},\\
	\chi_3 &=& \frac{\partial^3 m}{\partial H^3},
	\label{eq:def-chi3}
\end{eqnarray}
respectively. 
Here, $H$ is an external magnetic field [see Eq.~(\ref{eq:Zeeman})],
and $m$ is the magnetization per spin along the magnetic field.
Note that the susceptibilities are isotropic and independent of the direction of the magnetic field.

In the following MC simulations, we compute these susceptibilities by averaging the fluctuations at $H=0$ over the $x$, $y$, and $z$ directions as
\begin{eqnarray}
  \c &=& \sum_{\mu=x,y,z}\frac{\beta\Ns}{3} \left(\langle m_\mu^2 \rangle -\langle m_\mu\rangle^2\right) \nonumber\\
     &=& \sum_{\mu=x,y,z} \frac{\beta\Ns}{3}\langle m_\mu^2\rangle,\label{eq:chi_def}\\
          \nc &=& \sum_{\mu=x,y,z} \frac{\beta^3\Ns^3}{3} (\langle m_\mu^4 \rangle - 4 \langle m_\mu \rangle \langle m_\mu^3 \rangle - 3\langle m_\mu^2 \rangle^2\nonumber\\ 
    && +12 \langle m_\mu \rangle^2 \langle m_\mu^2 \rangle - 6\langle m_\mu \rangle ^4)\nonumber\\ 
    &=& \sum_{\mu=x,y,z}\frac{\beta^3 \Ns^{3}}{3}(\langle m_\mu^4 \rangle - 3\langle m_\mu^2 \rangle^2),
    \label{eq:MC-nc} 
\end{eqnarray}
where the magnetization in the $\mu$ direction ($\mu=x,y,z$) is defined by
\begin{eqnarray}
  m_\mu &=& \frac{1}{\Ns} \sum_i S_{i\mu},\label{eq:m}
\end{eqnarray}
and $\beta=1/T$ is the inverse temperature.
Note that $\langle m_\mu \rangle = 0$ for all the states considered in the present study.

We also compute the cubic susceptibility $\ncb$ defined by~\cite{Kobler96}
\begin{eqnarray}
        \ncb &=& 6\left(\frac{\partial^3 H}{\partial m^3}\right)^{-1}=-\frac{6\chi^4}{\nc}.\label{eq:def-cubic-chi3}
\end{eqnarray}

The sublattice magnetization $m_\text{s}$ is defined as 
\begin{eqnarray}
	m_{\text{s}}&\equiv& \frac{2}{\Ns} \left( \left \langle \left\langle \sum_{l}\left|\sum_{i \in l} \vec{S}_i\right|^2 \right\rangle_T\right\rangle_\Delta \right)^{1/2},
\end{eqnarray}
where $l$ labels the four sublattices of the pyrochlore lattice.
The specific heat $C$ is calculated by 
\begin{eqnarray}
	C&=&\frac{\left\langle \left\langle \mathcal{H}^2 \right\rangle_T - \left\langle \mathcal{H} \right\rangle_T^2 \right\rangle_\Delta}{\Ns T}.
\end{eqnarray}

All data shown in the following sections are averaged over a number of interaction sets varying from $100$ to $2000$. 
Typical MC steps for thermalization vary from $10^4$ to $10^7$ depending on $L$ and $\Delta$.
Monte Carlo steps for measurement are taken to be several times longer than those for thermalization.
Data obtained in independent MC runs for different interaction sets are splitted into several bins (typically 16).
Error bars are estimated by computing standard deviation for the bins.

%% file: phase_diagram.tex
\section{Phase diagrams and nature of phase transitions}\label{sec:phase-diagram}
In the following, we present the results for the model in Eq.~(\ref{eq:Ham2}) obtained by MC calculations.
Although some parts of the results have been already published in our previous paper,~\cite{Shinaoka10b}
we include them for making this paper self-contained and also for discussing the results in more comprehensive way.
In Sec.~\ref{sec:pd-overview}, we overview the phase diagrams obtained by MC simulations.
We show that the spin-lattice coupling induces peculiar SG behavior. 
A qualitative argument on its origin is given.
In Sec.~\ref{sec:pd-J2-off}, we focus on the case without $\Jcoop$.
We discuss the nature of the nematic and SG transitions for $J_2^\mathrm{coop}=0$ by showing MC data of the specific heat $C$, the spin collinearity $Q^2$, and the SG susceptibility $\chiSG$.
The results for $\Jcoop= 0.075$ are discussed in Sec.~\ref{sec:pd-J2-on}.
We show the data of the sublattice magnetization $m_\text{s}$, in addition to the above three quantities.
We discuss the nature of the spin-lattice order induced by $\Jcoop$ as well as effects of $\Jcoop$ on the SG behavior.

\subsection{Overview of calculated phase diagrams}\label{sec:pd-overview}
\begin{figure}[h]
 \centering
 \includegraphics[width=.4\textwidth,clip]{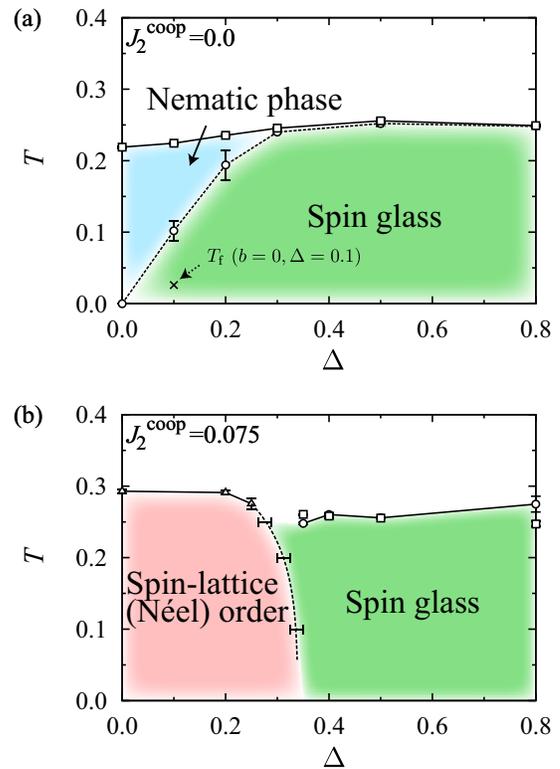}
 \caption{(Color online)
 $\Delta$-$T$ phase diagrams obtained by MC simulation at $b=0.2$: (a) $J_2^\mathrm{coop}=0.0$ and (b) $J_2^\mathrm{coop}=0.075$.
 The nematic ($\Tc$), antiferromagnetic ($\TN$), and SG transition temperatures ($\TSG$) are denoted by squares, triangles, and circles, respectively.
 In (a), $\TSG$ coincides with $\Tc$ for $\Delta \gtrsim 0.3$, suggesting a multicritical point at $\Delta \simeq 0.3~(\simeq b)$.
 The cross in (a) denotes $\TSG$ for $b=0$ and $\Delta=0.1$.~\cite{Saunders07,Andreanov10}
 See the text for details.
 }
 \label{fig:pd}
\end{figure}

Figures~\ref{fig:pd}(a) and \ref{fig:pd}(b) show the phase diagrams obtained at $J_2^\mathrm{coop}=0$ and $0.075$, respectively, for $b=0.2$.
These two cases give the answers for the questions in Figs.~\ref{fig:model-pd-schematic}(b) and \ref{fig:model-pd-schematic}(c), respectively.
We start with the results for $J_2^\mathrm{coop}=0$.
In the small $\Delta$ region ($\Delta \lesssim b$), as $T$ is lowered, the system undergoes successive two transitions:
a first-order nematic transition at $\Tc\simeq b$ and a second-order SG transition at $\TSG\propto \Delta$.
We call this regime the \textit{linear regime} because $\TSG$ grows approximately linearly with $\Delta$.
A remarkable observation is that $\TSG$ is largely enhanced compared to that in the bilinear limit ($b=0$)~\cite{Saunders07, Andreanov10};
the enhancement factor reaches about 3--5. 

At $\Delta\simeq b$, $\TSG$ appears to merge into $\Tc$. 
For larger $\Delta$, $\TSG~(=\Tc)$ becomes nearly independent of $\Delta$ and $\TSG \simeq b$,
which we call the \textit{plateau regime}.
This is in sharp contrast to the previously-reported SG behavior in the absence of the spin-lattice coupling, $\TSG \propto \Delta$.~\cite{Saunders07, Andreanov10}

\begin{figure}[h]
 \centering
 \includegraphics[width=.475\textwidth,clip]{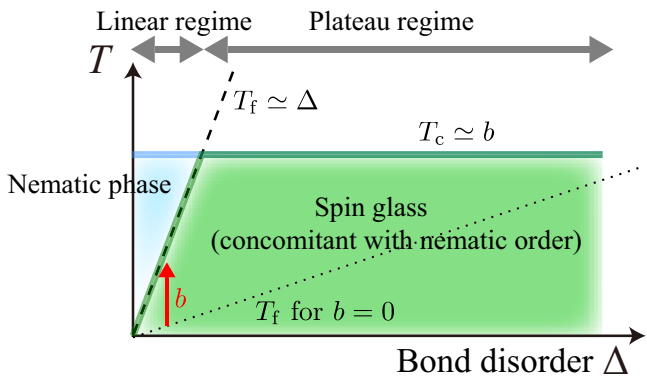}
 \caption{(Color online) Schematic $\Delta$-$T$ phase diagram for $b>0$ and $J_2^\mathrm{coop}=0$ [see Fig.~\ref{fig:pd}(a)].
 The nematic transition temperature $\Tc$ is almost independent of $\Delta$ as $\Tc\simeq b$.
 On the other hand, the SG transition temperature $\TSG$ increases linearly with $\Delta$ in the small $\Delta$ region.
 $\TSG$ is enhanced by $b$ compared to that in the bilinear limit ($b=0$) denoted by the dotted line.
 The nematic and SG transitions merge into a concomitant transition for $\Delta\gtrsim b$.
 }
 \label{fig:pd-schematic}
\end{figure}

Now, we give a qualitative description of the origin of the two peculiar aspects of the SG bevhavior:
(i) the enhancement of $\TSG$ by $b$ and (ii) the plateau behavior of $\TSG$ at $\TSG\simeq b$.
Figure~\ref{fig:pd-schematic} shows a schematic phase diagram for $0<b<J$ and $J_2^\mathrm{coop}=0$.
In the presence of the spin-lattice coupling $b$, the spin collinearity growing in the nematic phase below $T_c$ enforces spins to satisfy the spin-ice type local constraints, leading to the formation of locally-correlated collinear objects.
There, the system bears a semidiscrete degenerate manifold with multivalley energy landscape as illustrated in Fig.~\ref{fig:multivalley}(a).
This strongly suppresses thermal fluctuations compared to the bilinear case with $b=0$ where the degenerate manifold is continuously connected.
At the same time, the spin-spin correlations are much enhanced to exhibit quasi-long-range behavior below $\Tc$ due to the spin-ice type macroscopic degeneracy.~\cite{Shannon10}
As illustrated in Fig.~\ref{fig:pd-schematic}, these effects enhance $\TSG$ from the dotted line of $\TSG$ for $b=0$ to the broken line $\TSG\simeq\Delta$.
This mechanism, however, does not work above $\Tc$.
As a result, while increasing $\Delta$, $\TSG$ is saturated at $\Tc\simeq b$, leading to the plateau behavior of $\TSG$ for $\Delta\gtrsim b$.
Note that the plateau behavior of $\TSG$ is transient;
namely, $\TSG$ will increase again for a sufficiently large $\Delta$, presumably along the extension of the dotted line of $\TSG$ for $b=0$.
(Such behavior is not observed for the current parameter sets.)

Let us move onto the results for $J_2^\mathrm{coop}=0.075$ [Fig.~\ref{fig:pd}(b)].
In this case, the nematic phase is taken over completely by the $\vec{q}=0$ spin-lattice (N\'{e}el) order,
whose transition temperature $\TN$ is in the energy scale of $J_2^\mathrm{coop}$.
As $\Delta$ increases, the spin-lattice order vanishes around $\Delta\simeq \TN$.
For larger $\Delta$, a concomitant transition of nematic and SG is seen at $\Tc=\TSG\simeq b$ similarly to the case with the cooperative coupling.
Furthermore, as discussed later on, the thermodynamic properties in the plateau regime are essentially the same as at $J_2^\mathrm{coop}=0$.
These indicate that the plateau behavior of $\TSG$ is robust against the cooperative aspect of bond distortions.


\subsection{Case without the cooperative coupling: $J_2^\mathrm{coop}=0$}\label{sec:pd-J2-off}
\begin{figure*}
 \centering
 \includegraphics[width=0.85\textwidth,clip]{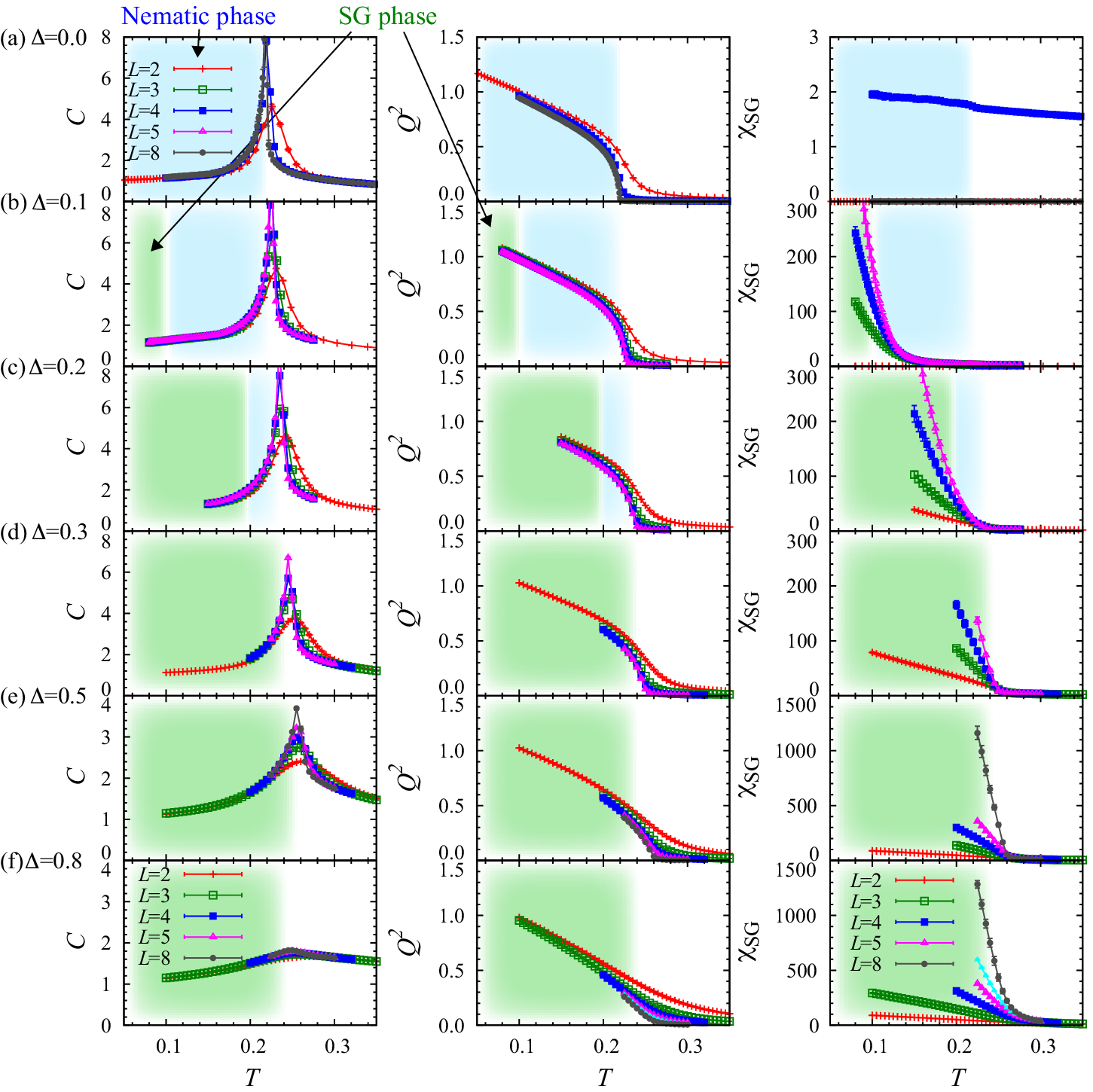}
 \caption{(Color online) The specific heat $C$, spin collinearity $Q^2$, and SG susceptibility $\chiSG$ calculated at $b=0.2$ for (a) $\Delta=0.0$, (b) $0.1$, (c) $0.2$, (d) $0.3$, (e) $0.5$, and (f) $0.8$.
 The data are calculated for the system sizes ranging from $L=2$ (128 spins) to $L=8$ (8192 spins).
}
 \label{fig:MC-data-all}
\end{figure*}
\subsubsection{Successive nematic and spin-glass transitions in the linear regime}
\begin{figure}[h]
 \centering
 \includegraphics[width=0.35\textwidth,clip]{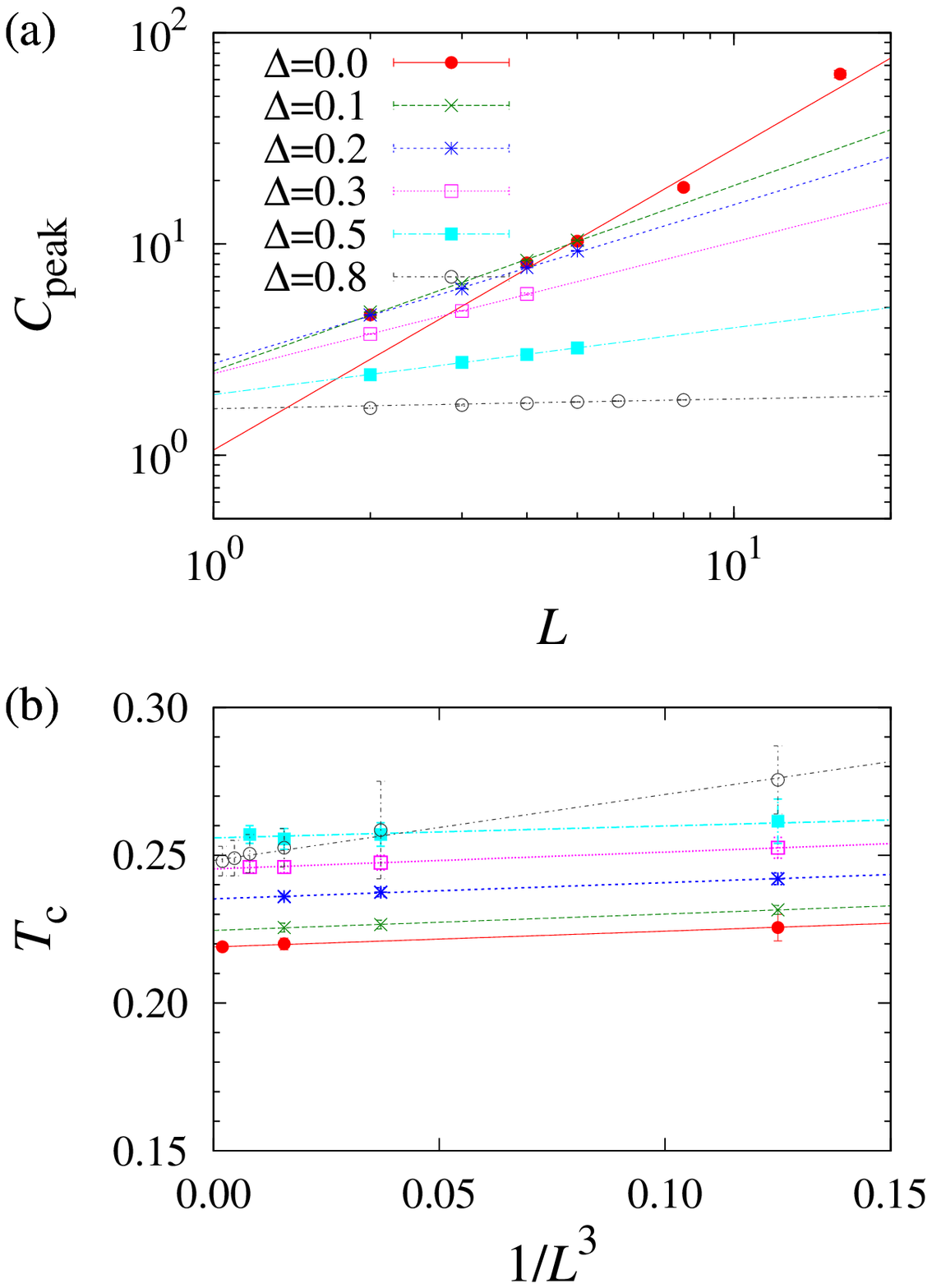}
 \caption{(Color online) 
 (a) $L$ dependence of the peak values of the specific heat $C$, $C_\mathrm{peak}$. The lines are power-law fitting by $C_\mathrm{peak}\propto L^p$.
 The obtained values of $p$ are shown in Table~\ref{table:fss-linear}.
 (b) System size dependences of the peak temperature of $C$. The lines represent the extrapolations of $C_\mathrm{peak}$ to the bulk limit with $\Tc(L) - \Tc \propto 1/L^3$.
 The obtained values of $\Tc$ are summarized in Table~\ref{table:fss-linear}.
 }
 \label{fig:Cpeak}
\end{figure}

We discuss the nature of the successive nematic and SG transitions in the linear regime in the case of $J_2^\mathrm{coop}=0$.
Figures~\ref{fig:MC-data-all}(a)-(d) show the specific heat $C$, spin collinearity $Q^2$, and SG susceptibility $\chiSG$ calculated for $\Delta \lesssim 0.3$.
One can clearly see that $C$ exhibits a sharp peak concurrently with the onset of $Q^2$ at $\Tc \simeq 0.2$--$0.25$.
These indicate the nematic transition.
Furthermore, the peak value of $C$, $C_\mathrm{peak}$, appears to diverge in the thermodynamic limit; 
the data are well fitted by 
\begin{eqnarray}
  C_\mathrm{peak}&\propto& L^p\label{eq:Cpeak-p}
\end{eqnarray}
with $p>0$ as shown in Fig.~\ref{fig:Cpeak}(a) (see also Table~\ref{table:fss-linear}).
We estimated $\Tc$ by extrapolating the peak temperature of $C$ to the thermodynamic limit, as shown in Fig.~\ref{fig:Cpeak}(b).
The resulting values of $\Tc$ are summarized in Table~\ref{table:fss-linear} and plotted in Fig.~\ref{fig:pd}(a).
It is noteworthy that $\Tc$ is almost independent of $\Delta$ or even enhanced by $\Delta$ slightly.
This is presumably because of the competition between the randomness in $b_{ij}$ and $J_{ij}$:
The former suppresses local spin collinearity, while the latter does the opposite.~\cite{Bellier-Castella01}

\begin{figure}
 \centering
 \includegraphics[width=.35\textwidth,clip]{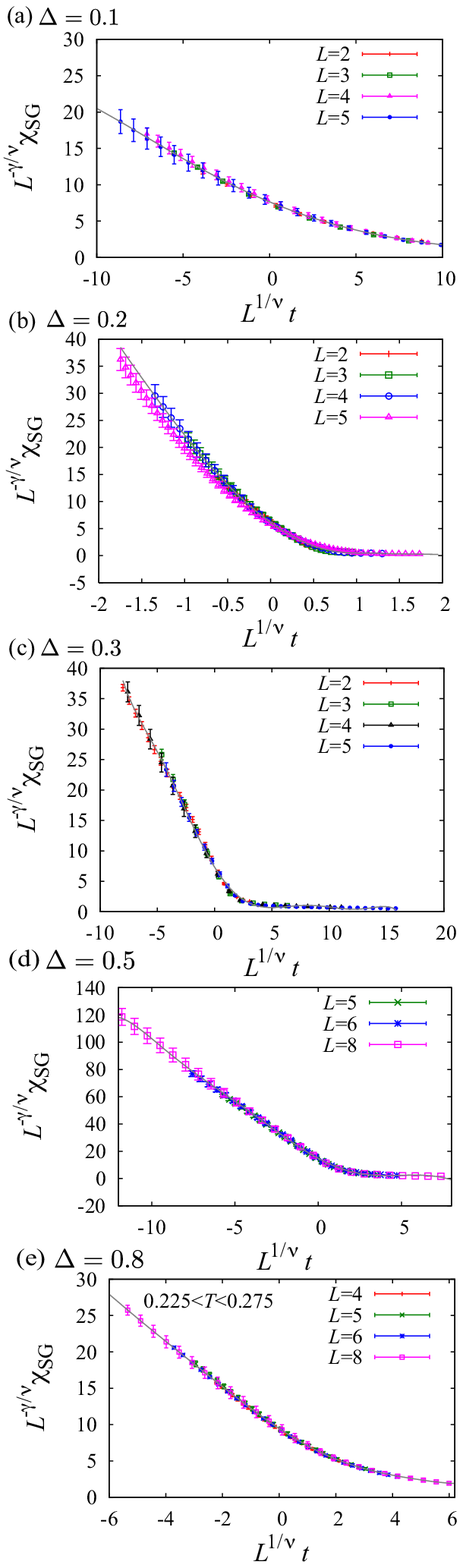}
 \caption{(Color online) Scaling collapse of $\chiSG$ calculated with $\Jcoop=0$ at (a)
 $\Delta=0.1$, (b) 0.2, (c) 0.3, (d) 0.5, and (e) 0.8.
 The estimated $\TSG$ and the critical exponents are shown in Table~\ref{table:fss-linear}.
 }
 \label{fig:scaling-collapse-linear}
\end{figure}

On the other hand, $\chiSG$ shows divergent behavior at a lower $T$, as shown in Fig.~\ref{fig:MC-data-all}.
This is a signature of the SG transition.
In order to estimate the SG transition temperature $\TSG$, we perform the finite-size scaling analysis by assuming
\begin{eqnarray}
	\chiSG &=& L^{\gamma/\nu} f(L^{1/\nu}t).\label{eq:scaling-chiSG}
\end{eqnarray}
Here $t=(T-\TSG)/\TSG$, $\nu$ and $\gamma$ are the critical exponents for the correlation length and $\chiSG$, respectively.
Figures~\ref{fig:scaling-collapse-linear}(a)-\ref{fig:scaling-collapse-linear}(c) show the scaling collapses of MC data obtained at $\Delta=0.1$, 0.2, and 0.3, respectively.
All the MC data collapse onto a single curve within error bars throughout the linear regime.
The resulting $\TSG$ and the critical exponents are listed in Table~\ref{table:fss-linear}.
The values of the critical exponents for $\Delta=0.1$ are consistent with those in the bilinear limit $b=0$~\cite{Saunders07, Andreanov10} 
as well as of the canonical SG~\cite{Nakamura02,Fernandez09} within the error bars. 
Note that $\gamma$ becomes smaller as approaching the multicritical point near $\Delta=0.3$.
Also, the scaling for $\Delta=0.2$ in Fig.~\ref{fig:scaling-collapse-linear}(b) show rather poor convergence. 
These are presumably due to finite-size effects, which become more conspicuous when $\TSG$ comes close to $T_c$.
\begin{table}[h]
 \centering
 \begin{tabular}{c|cc|ccc}\hline
	 & \multicolumn{2}{c|}{Nematic transition}  &\multicolumn{3}{c}{SG transition}    \\ \hline
	 $\Delta$ & $\Tc$       &     $p$    &$\TSG$      & $\gamma$ & $\nu$    \\ \hline
   $0.0$    & 0.219(1)    &   1.4(1)   &   --       &  --      &  --      \\ 
   $0.1$    & 0.225(1)    &   0.88(4)  & 0.102(14)  & 2.24(75) & 1.16(18) \\ 
   $0.2$    & 0.236(1)    &   0.75(3)  & 0.20(2)    & 2(1)    & 0.9(3)   \\
   $0.3$    & 0.246(1)    &   0.623(5) & 0.240(2)   & 0.6(2)   & 0.58(6)  \\
   $0.5$    & 0.256(1)    &   0.317(6) & 0.256(1)   & 0.71(6)  & 0.65(1)  \\
   $0.8$    & 0.2482(4)   &   0.046(2) & 0.248(2)   & 1.5(1)   & 0.80(2)  \\ \hline
 \end{tabular}
 \caption{Transition temperatures and critical exponents for the nematic and SG transitions.
 We estimated $\Tc$ and $p$ by the finite-size analysis of $C$ (see Fig.~\ref{fig:Cpeak}).
 The SG transition temperatures $\TSG$ and the exponents $\gamma$ and $\nu$ are estimated by finite-size scaling of $\chiSG$ (see Fig.~\ref{fig:scaling-collapse-linear}).
 } 
 \label{table:fss-linear}
\end{table}

\begin{figure}
 \centering
 \includegraphics[width=.4\textwidth,clip]{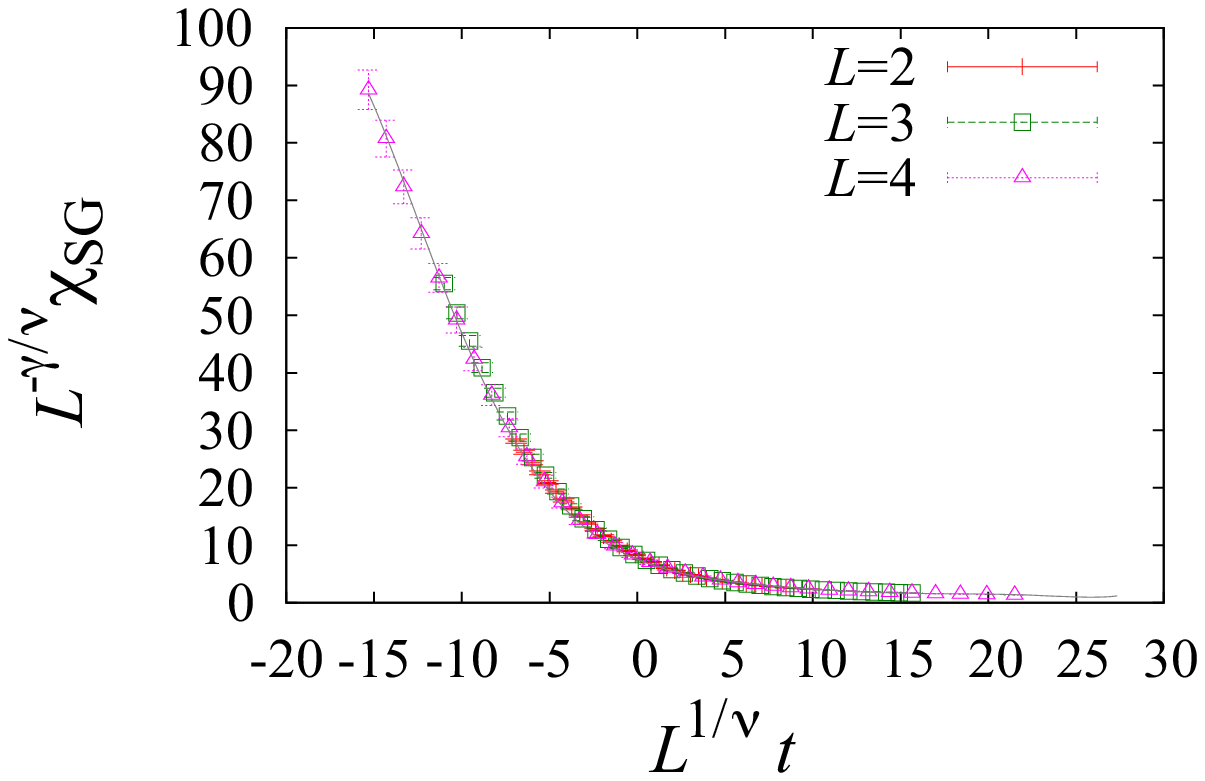}
 \caption{(Color online) Scaling collapse of $\chiSG$ calculated for the Ising model given in Eq.~(\ref{eq:Ham-Ising}).
 We obtained $\TSG=0.151(2)$, $\gamma=0.79(2)$, and $\nu=0.86(2)$ for $L=2,3$, and $4$.}
 \label{fig:ising-scaling}
\end{figure}

Now, we examine the effect of spin collinearity induced by the spin-lattice coupling $b$ on $\TSG$.
As listed in Table~\ref{table:fss-linear}, the estimated $\TSG$ in the linear regime is largely enhanced from the value in the bilinear limit $b$; e.g., for $\Delta=0.1$, $\TSG=0.102(14)$ at $b=0.2$, which is $3$--$5$ times larger than $\TSG=0.02$-$0.032$ at $b=0$.~\cite{Saunders07, Andreanov10}
In order to clarify the behavior in the collinear limit $b \to \infty$, we consider an Ising counterpart of the present model:
\begin{equation}
	\mathcal{H}= \sum_{\langle i,j \rangle} J_{ij} \sigma_i\sigma_j.\label{eq:Ham-Ising}
\end{equation}
Here, $\sigma_i~(=\pm 1)$ denotes an Ising spin at site $i$,
and $J_{ij}$ are the bond-disordered antiferromagnetic exchange interactions defined in Eq.~(\ref{eq:Jij}).
The ground state has spin-ice type macroscopic degeneracy with discrete energy landscape.
Figure~\ref{fig:ising-scaling} shows a scaling collapse of $\chiSG$ calculated for the model (\ref{eq:Ham-Ising}) at $\Delta=0.1$.
We obtained $\TSG=0.151(2)$, which is 5--8 times higher than that in the bilinear Heisenberg limit $b=0$.~\cite{Saunders07, Andreanov10}
This indicates that the discrete structure of the degenerate manifold enhances $\TSG$.
The result supports that the spin collinearity and associated semidiscrete manifold emergent below $\Tc$ can be responsible for the remarkable enhancement of $\TSG$ by $b$.
Also, it suggests that the enhancement factor of $\TSG$ ranges up to 5--8 at $\Delta=0.1$ depending on the value of $b$.

\subsubsection{Concomitant transition in the plateau regime}\label{sec:plateau}
The two successive transitions merge into a single transition at $\Delta\simeq b$; that is, for larger $\Delta$, $\chiSG$ diverges concurrently with the onset of $Q^2$, as shown in Fig.~\ref{fig:MC-data-all}.
We estimated $\Tc$ by extrapolating the peak temperature of $C$ in the same manner as in the linear regime [see Fig.~\ref{fig:Cpeak}(b) and Table~\ref{table:fss-linear}].
We also performed the finite-size analysis for $\chiSG$ to estimate $\TSG$; we successfully obtained scaling collapses, as shown in Figs.~\ref{fig:scaling-collapse-linear}(d) and \ref{fig:scaling-collapse-linear}(e).
As shown in Table~\ref{table:fss-linear}, $\Tc$ and $\TSG$ estimated independently coincide with each other within error bars in the plateau regime $\Delta \gtrsim 0.3$, 
indicating that the nematic and SG transitions occur concomitantly. 
The results also indicate that the MC data are compatible with the second-order transition, in contrast to the weak first-order transition at $T_c$ in the linear regime.

\begin{figure}
 \centering
 \includegraphics[width=0.25\textwidth,clip]{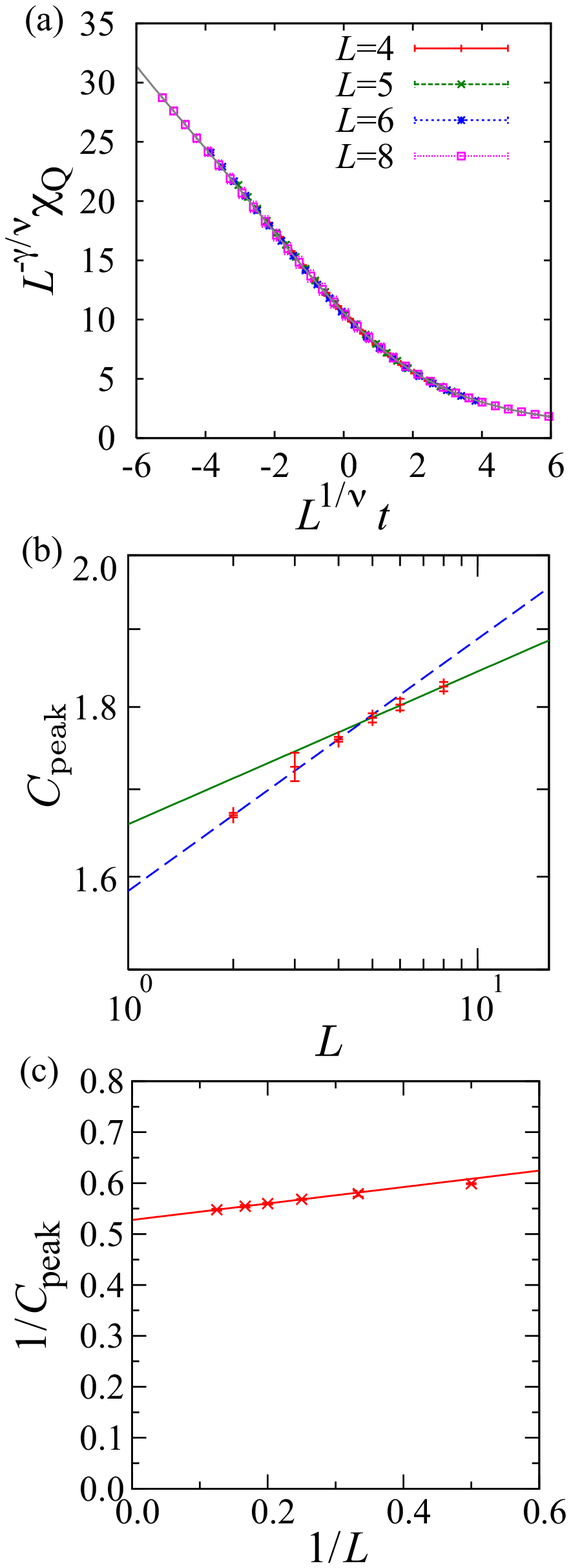}
 \caption{(Color online) 
 Finite-size scaling analysis of the nematic transition at $\Delta=0.8$ and $\Jcoop=0$.
 (a) Scaling collapse of $\chi_Q$ for the data in the range of $0.225\le T\le 0.275$. We obtained $\TSG=0.248(2)$, $\gamma_Q=1.5(1)$, and $\nu_Q=0.80(2)$.
 (b) Semi-logarithmic plot of the $L$ dependence of $C_\mathrm{peak}$.
 (c) $1/C_\mathrm{peak}$ as a function of $1/L$.
 }
 \label{fig:fss-Q}
\end{figure}

To examine the critical properties of the concomitant transition in more detail,
we perform a finite-size scaling analysis for $\chi_Q$ at $\Delta=0.8$.
Similarly to $\chiSG$, we assume
\begin{eqnarray}
	\chi_Q &=& L^{\gamma_{Q}/\nu_Q} f_{Q}(L^{1/\nu_Q}t),
\end{eqnarray}
where $t=(T-T_\mathrm{c})/T_\mathrm{c}$, and $\nu_Q$ and $\gamma_Q$ are the critical exponents for the correlation length and $\chi_Q$, respectively.
As demonstrated in Fig.~\ref{fig:fss-Q}(a),
we successfully obtained a scaling collapse of the data for $4\le L\le 8$ with $\Tc=0.249(1)$, $\gamma_Q=1.6(1)$, and $\nu_Q=0.78(2)$.
It is worthy noting that we observed no significant system-size dependence in the scaling results as shown in Table~\ref{table:finite-size-effect};
the estimates for different range of $L$, i.e., $2\le L\le 4$ and $4\le L\le 8$, coincide with each other within the error bars. 
The value of $\Tc$ is consistent with that estimated by the extrapolation of the peak temperature of $C$.
Furthermore, $\Tc$ and the critical exponents are consistent with 
those obtained by the finite-size scaling analysis of $\chiSG$ within error bars (see Table~\ref{table:fss-linear}).
All of these results 
provide strong evidence for the concomitant nature of the SG and nematic transitions;
two transitions occur concomitantly, in a second-order fashion with the identical critical exponents.

As seen in Figs.~\ref{fig:MC-data-all}(e) and \ref{fig:MC-data-all}(f),
the peak in $C$ is markedly suppressed and broadened in the plateau regime.
Let us focus on the result at $\Delta=0.8$ in Fig.~\ref{fig:MC-data-all}(f).
As shown in Fig.~\ref{fig:fss-Q}(b),
the peak value $C_\mathrm{peak}$ shows a very weak $L$ dependence. 
The growth gets slower as $L$ increases; when we fit the data by $C_\mathrm{peak}\propto L^\alpha$, the exponent $\alpha$ decreases as $L$ increases [$\alpha=0.076(1)$ for $2\le L\le 4$, and $\alpha=0.046(2)$ for $5\le L\le 8$].
Alternatively, the data can be well fitted by assuming $C_\mathrm{peak}^{-1}=aL^q+C_\mathrm{peak}^{-1}(\infty)$ with $q \simeq -1$ [see Fig.~\ref{fig:fss-Q}(c)].
These results suggest that $C$ is non-singular in the thermodynamic limit.
The broad peak behavior is apparently similar to that observed in the canonical SG,~\cite{Binder86}
but the peak is located at $\TSG~(=\Tc)$ in the present case.
This is in contrast to the case of the canonical SG in which the peak temperature $T_\mathrm{peak}$ exceeds $\TSG$ typically by 20\%.~\cite{Binder86}
The broad peak at $\TSG~(=\Tc)$ will be of characteristic of the SG transition concomitant with the nematic transition in the present system.
We will discuss effects of a magnetic field on the peak structure in Sec.~\ref{sec:mag-field}.
Comparisons with experiments are given in Sec.~\ref{sec:comparison-with-exp}.
\begin{table}
 \centering
 \begin{tabular}{l|lll|lll}\hline
	 System sizes & \multicolumn{3}{c|}{$\chiSG$}      & \multicolumn{3}{c}{$\chi_Q$}  \\ \cline{2-7}
               & $\TSG$       & $\gamma$ & $\nu$   & $\Tc$    & $\gamma_{Q}$ & $\nu_Q$   \\ \hline
   $L=2,3,4$   &  0.248(2)    & 1.57(7)  & 0.79(2) & 0.242(2) & 2.0(3)   & 0.87(7) \\
   $L=4,5,6,8$ &  0.248(2)    & 1.5(1)   & 0.80(2) & 0.249(1) & 1.6(1)   & 0.78(2) \\ \hline
 \end{tabular}
 \caption{
 Comparison of $\TSG$, $T_c$, and critical exponents at $\Delta=0.8$ and $\Jcoop=0.8$ for different sets of $L$ used in the finite-size scaling analysis. 
 The upper row shows the results obtained for the set of $L=2,3,4$, while the lower for $L=4,5,6,8$.
 }
 \label{table:finite-size-effect}
\end{table}

\subsection{Case with the cooperative coupling: $J_2^\mathrm{coop}>0$}\label{sec:pd-J2-on}
\begin{figure*}
 \centering
 \includegraphics[width=0.85\textwidth,clip]{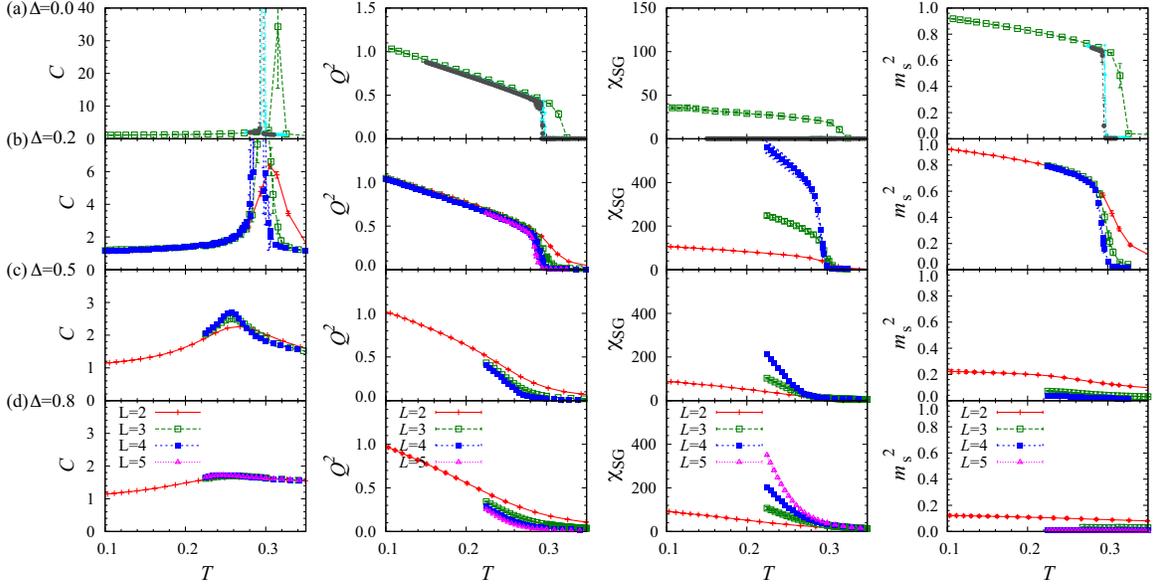}
 \caption{(Color online) The specific heat $C$, spin collinearity $Q^2$, SG susceptibility $\chiSG$, and the square of sublattice magnetization, $m_\mathrm{s}^2$, calculated at $b=0.2$ and $J_2^\mathrm{coop}=0.075$: (a) $\Delta=0.0$, (b) 0.2, (c) 0.5, and (d) 0.8. 
 }
 \label{fig:MC-data-all-J2}
\end{figure*}

\begin{figure}[h]
 \centering
 \includegraphics[width=.4\textwidth,clip]{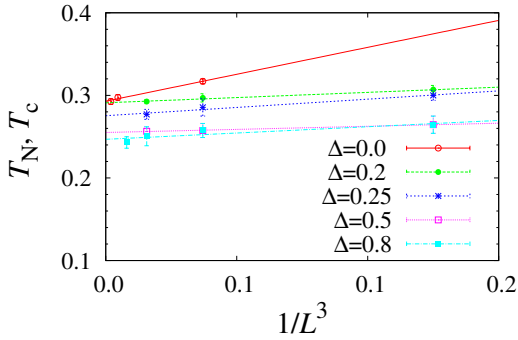}
 \caption{(Color online) Extrapolation of the peak temperatures of the specific heat $C$ to the bulk limit.
 The data are calculated with $b=0.2$ and $\Jcoop=0.075$.}
 \label{fig:Cpeak-J2}
\end{figure}

Now, we move onto the results with the cooperative coupling $J_2^\mathrm{coop}=0.075$.
In the small $\Delta$ region ($\Delta\lesssim 0.3$), the system undergoes a first-order transition to
the spin-lattice (N\'{e}el) ordered state at $\TN$.
Typical MC data calculated at $\Delta=0$ and 0.2 are shown in Figs.~\ref{fig:MC-data-all-J2}(a) and \ref{fig:MC-data-all-J2}(b), respectively.
The square of sublattice magnetization $m_\mathrm{s}^2$ exhibits a steep rise at $\TN\simeq 0.3$, and the specific heat $C$ shows a sharp peak at the same time.
These clearly indicate that the transition is first order.
We estimated the N\'{e}el transition temperature $\TN$ by extrapolating the peak temperature of $C$ to the bulk limit (see Fig.~\ref{fig:Cpeak-J2}).
The obtained values are $\TN= 0.293(3)$, 0.291(3), and 0.275(7) for $\Delta=0$, 0.2, and 0.25, respectively; the values are plotted in Fig.~\ref{fig:pd}(b).

\begin{figure}[h]
 \centering
 \includegraphics[width=.4\textwidth,clip]{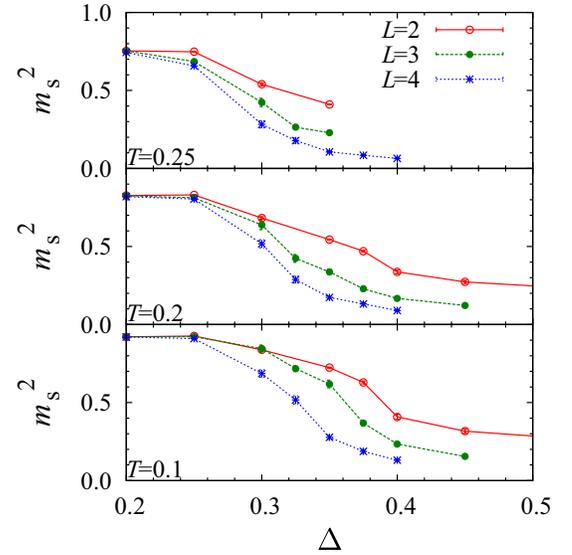}
 \caption{(Color online) Squared sublattice magnetization $m_{\text{s}}^2$ as a function of $\Delta$ at (a) $T=0.25$, (b) 0.2, and (c) 0.1. We take $b=0.2$ and $\Jcoop=0.075$.}
 \label{fig:sbl-mag2}
\end{figure}

As $\Delta$ increases, the spin-lattice ordered phase is destabilized by disorder; 
e.g., at $\Delta=0.5$, $m_\mathrm{s}^2$ decreases as $L$ increases even at the lowest $T~(\simeq 0.22)$ investigated, as shown in Fig.~\ref{fig:MC-data-all-J2}(c).
To estimate the critical value of $\Delta$, we plot $m_\mathrm{s}^2$ as functions of $\Delta$ at $T=0.25$, 0.2, and 0.1 in Fig.~\ref{fig:sbl-mag2}.
We estimated the phase boundary by the inflection point of $m_{\text{s}}^2(\Delta)$ curve at each $T$ and plotted them in Fig.~\ref{fig:pd}(b).

\begin{figure}[h]
 \centering
 \includegraphics[width=0.475\textwidth,clip]{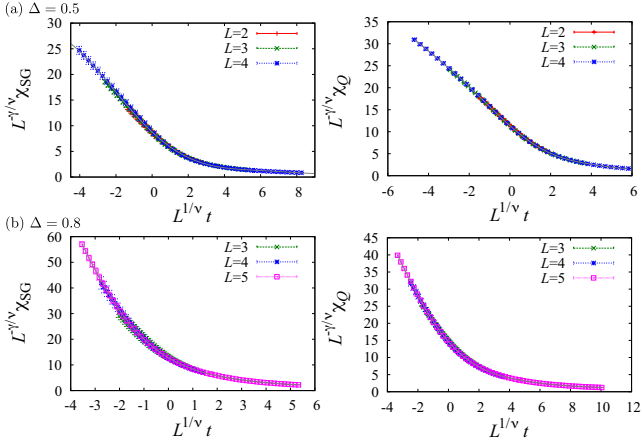}
 \caption{(Color online) Scaling collapses of the SG susceptibility $\chiSG$ and nematic susceptibility $\chi_Q$ at (a) $\Delta=0.5$ and (b) 0.8 for $J_2^\mathrm{coop}=0.075$ and $b=0.2$ in the plateau regime [see Fig.~\ref{fig:pd}(b)].
 We take $b=0.2$ and $\Jcoop=0.075$.
 The estimated values for the transition temperatures and the critical exponents are presented in Table~\ref{table:finite-size-scaling-J2}.
 }
 \label{fig:scaling-collapse-J2}
\end{figure}

For larger $\Delta\gtrsim 0.3$, the system exhibits a single and concomitant transition of SG and nematic at $T\simeq b$
similarly to that in  the plateau regime for $J_2^\mathrm{coop}=0$.
Typical MC data in this regime are shown in Figs.~\ref{fig:MC-data-all-J2}(c) and \ref{fig:MC-data-all-J2}(d).
At $T\simeq b$, $Q^2$ shows a rapid increase, which is accompanied by a broad peak in $C$.
Below the same $T$, $\chiSG$ shows divergent behavior as $L$ increases.
As shown in Fig.~\ref{fig:Cpeak-J2}, we estimated $\Tc$ by extrapolating the peak temperatures of $C$ to the bulk limit.
On the other hand, we estimate the transition temperatures by the finite-size scaling of $\chiSG$ and $\chi_Q$ as in the case of $J_2^\mathrm{coop}=0$;
we successfully obtained scaling collapses for $0.35 \le \Delta \le 0.8$.
The typical results obtained for $\Delta=0.5$ and 0.8 are shown in Fig.~\ref{fig:scaling-collapse-J2}.
The values of $\Tc$ and $\TSG$ as well as the critical exponents are shown in Table~\ref{table:finite-size-scaling-J2}.
The values for the exponents obtained for $\Delta=0.8$ are consistent with those for $J_2^\mathrm{coop}=0$ listed in Tables~\ref{table:fss-linear} and \ref{table:finite-size-effect}.
This indicates that the critical properties of the concomitant transitions in the plateau regime are essentially the same for $J_2^\mathrm{coop}=0$ and 0.075.

\begin{table}[h]
 \centering
 \begin{tabular}{l|lll|lll}\hline
	 $\Delta$    & \multicolumn{3}{c|}{$\chiSG$}      & \multicolumn{3}{c}{$\chi_Q$}  \\ \cline{2-7}
               & $\TSG$       & $\gamma$ & $\nu$   & $\Tc$    & $\gamma_Q$ & $\nu_Q$   \\ \hline
   $0.5$       &  0.256(1)    & 1.02(2)  & 0.65(1) & 0.258(1) & 1.15(3)   & 0.62(1) \\
   $0.8$       &  0.27(2)     & 1.1(4)   & 0.97(5) & 0.255(5) & 1.4(1)    & 0.83(2) \\ \hline
 \end{tabular}
 \caption{Comparison of the transition temperatures and critical exponents of the SG and nematic transitions for $b=0.2$ and $J_2^\mathrm{coop}=0.075$. 
 The values are estimated from the finite-size scaling in Fig.~\ref{fig:scaling-collapse-J2}.
 }
 \label{table:finite-size-scaling-J2}
\end{table}

%% file: magnetic_susceptibility.tex
\section{Magnetic susceptibility}\label{sec:susceptibility}
In this section, we investigate effects of the spin-lattice coupling on the magnetic susceptibility in the bilinear-biquadratic model.
In Sec.~\ref{sec:mag-chi1}, we discuss the linear susceptibility in the high-$T$ paramagnetic phase.
A difference between the FC and ZFC susceptibilities in the SG phase are investigated in Sec.~\ref{sec:fc}.
We analyze high-$T$ behavior and critical properties of nonlinear susceptibilities in Sec.~\ref{sec:chi3}.
Throughout these sections, we focus on the case without the cooperative coupling: $\Jcoop=0$.

\subsection{Linear susceptibility in the paramagnetic phase}\label{sec:mag-chi1}
\begin{figure}[h]
 \centering
 \includegraphics[width=0.4\textwidth,clip]{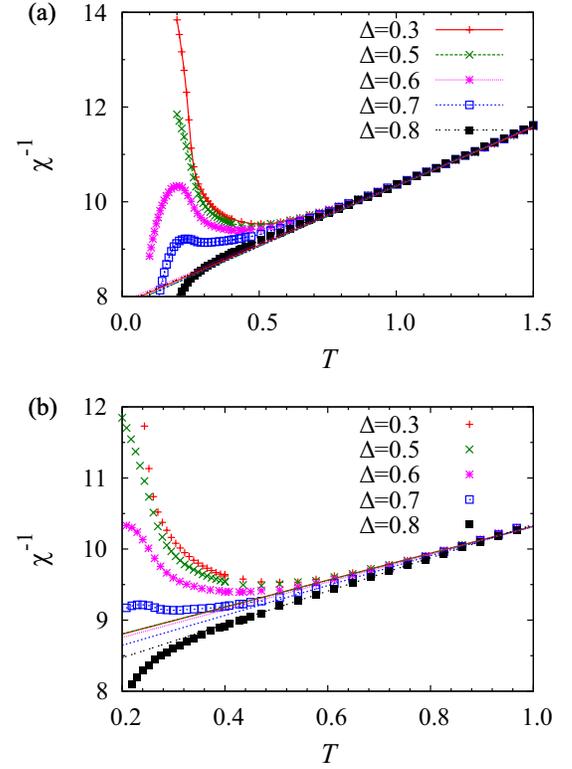}
 \caption{(Color online) $T$ dependence of the inverse of the magnetic susceptibility $\chi$ calculated at $b=0.2$ and $J_2^\mathrm{coop}=0$ for $L=3$.
 The data are plotted for $0.0\le T \le 1.5$ and $0.2\le T\le 1.0$ in (a) and (b), respectively.
 The lines in (a) and (b) denote the fits by the Curie-Weiss law in Eq.~(\ref{eq:CWlaw}) in the range of $1.2\le T\le 1.5$ and $0.6 \le T\le 0.9$, respectively.
 }
 \label{fig:chi1}
\end{figure}

Let us first discuss the $T$ dependence of susceptibility $\chi$ defined in Eq.~(\ref{eq:chi_def}). 
The result calculated at $b=0.2$ is shown in Fig.~\ref{fig:chi1}.
At high $T>1.0$, the data are well fitted by the Curie-Weiss law:
\begin{eqnarray}
	\chi = \frac{C_\mathrm{CW}}{T-\theta_\mathrm{CW}},
	\label{eq:CWlaw}
\end{eqnarray}
where $\theta_\mathrm{CW}$ is the Curie-Weiss temperature and $C_\mathrm{CW}$ is the Curie-Weiss constant.
From the fitting in the range of $1.0 < T < 1.5$, we obtain $\theta_\mathrm{CW}\simeq -3.1$ and $C_\mathrm{CW}=0.39$--$0.41$. 
The estimates show deviations from the expected values, $\theta_\mathrm{CW}=-4$ and $C_\mathrm{CW}=1/3$, for the present model with the mean value of $J_{ij}$ unity and $|\vec{S}_i|=1$.
The deviations is presumably because the $T$ range for the fitting is not high enough.

On the other hand, at lower $T$, the $T$ dependence of $\chi$ deviates from the Curie-Weiss law. 
In particular, below $T \sim 0.5$, $\chi$ is suppressed from the Curie-Weiss behavior for small $\Delta$, presumably due to the growth of antiferromagnetic correlations. 
Meanwhile, the low-$T$ part is increased as $\Delta$ increases.
This enhancement of $\chi$ may be ascribed to the existence of spins which are weakly coupled to their neighbors in the presence of randomness.

Figure~\ref{fig:chi1}(b) shows an enlarged plot of the $T$ dependence of $\chi$ for $T<1$.
Interestingly, $\chi$ shows Curie-Weiss-like $T$ dependence with different $\theta_\mathrm{CW}$ and $C_\mathrm{CW}$ in this intermediate-$T$ range.
Figure~\ref{fig:chi1}(b) shows the results of fitting in the range of $0.6\le T\le 0.9$.
We found that the estimated value of $\theta_\mathrm{CW}$ sensitively increases as $\Delta$ increases; for instance, $\theta_\mathrm{CW}$ increases from $-4.5$ at $\Delta=0.5$ to $-3.5$ at $\Delta=0.8$.
On the other hand, the estimated value of $C_\mathrm{CW}$ decreases as $\Delta$ increases; from $C_\mathrm{CW}=0.53$ at $\Delta=0.5$ to $0.43$ at $\Delta=0.8$.
Comparisons with experiments are given in Sec.~\ref{sec:comparison-with-exp}.

\subsection{Hysteresis in the susceptibility in the SG phase}\label{sec:fc}
Now, we discuss the hysteresis of magnetic susceptibility in the SG phase.
In the canonical SG, the magnetic susceptibility shows hysteresis below $\TSG$, i.e., different $T$ dependence between FC and ZFC susceptibilities. 
Such magnetic hysteresis was seen also in frustrated SG materials, e.g., Y$_2$Mo$_2$O$_7$~\cite{Gingras97} and CoAl$_2$O$_4$.~\cite{Hanashima13}

To compare the SG behavior in the present model with experiments, we compute the FC and ZFC susceptibilities by MC simulation as follows.
For the FC susceptibility, we first thermalize the system in the paramagnetic phase in an external magnetic field $H$ by adding the Zeeman term to the Hamiltonian:
\begin{eqnarray}
  \mathcal{H}_\mathrm{Zeeman} &=& - H \sum_i S_{iz}.
\label{eq:Zeeman}
\end{eqnarray}
Then, the system is cooled down in steps of $\Delta T=0.05$.
The system is equilibrated at each $T$ for 1000 MC steps, in which magnetization is measured simultaneously.
On the other hand, for the ZFC susceptibility, we cool down the system in a similar manner to the FC case but in the absense of magnetic field.
We store the spin configurations at each $T$ in the cooling processes.
Then, we apply a magnetic field to the system at each $T$ and measure the magnetization for 1000 MC steps.
In the simulations, we use only the single-spin update and overrelaxation.
We omit the loop algorithm, as such global relaxation process is presumably absent in real systems.

\begin{figure}[h]
 \centering
 \includegraphics[width=0.5\textwidth,clip]{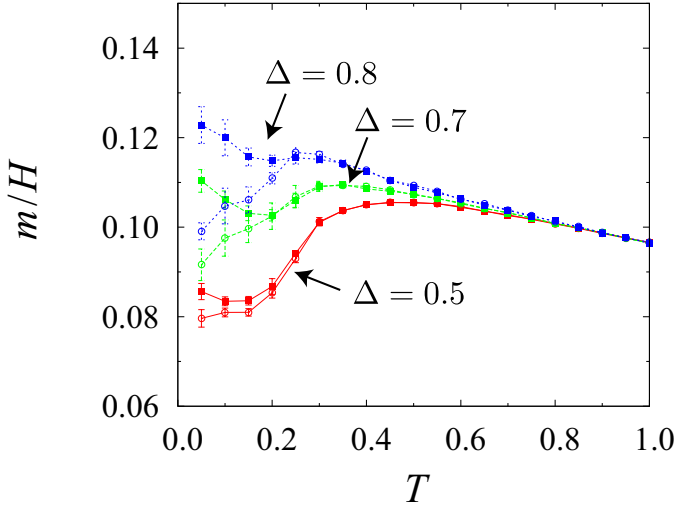}
 \caption{(Color online) 
   Field-cooled (FC, filled squares) and zero field-cooled (ZFC, open circles) susceptibilities calculated at $b=0.2$ and $\Jcoop=0$.
   The data are calculated at $H=0.1$ for $L=3$. 
 }
 \label{fig:chi-ZFC-FC}
\end{figure}

We show the results for the FC and ZFC susceptibilities in the plateau regime in Fig.~\ref{fig:chi-ZFC-FC}.
At $\Delta=0.5$, which is close to the multicritical point, the FC and ZFC susceptibilities are suppressed below $\TSG$ due to the spin collinearity induced by $b$.
At the same time, a difference appears between the FC and ZFC susceptibilities below $\TSG$, reflecting spin freezing.
As $\Delta$ increases, the suppression of the susceptibility below $\TSG$ becomes less pronounced, while the difference between the FC and ZFC data becomes more apparent.
In particular, at $\Delta=0.8$, the FC susceptibility increases continuously below the transition temperature, being in contrast to the result for $\Delta=0.5$.
Thus, our model reproduces the hysteresis behavior of the magnetic susceptibility observed in the frustrated SG materials.

\subsection{Nonlinear magnetic susceptibilities}\label{sec:chi3}
As seen in Sec.~\ref{sec:mag-chi1}, the spin-lattice coupling $b$ does not affect the high-$T$ behavior of the linear susceptibility.
Instead, a fingerprint of $b$ appears in the cubic susceptibility $\ncb$ defined by Eq.~(\ref{eq:def-cubic-chi3}): at high $T$, $\ncb$ obeys a Curie-Weiss-like law as 
\begin{eqnarray}
  \ncb  &=& 6\left(\frac{\partial^3 H}{\partial m^3}\right)^{-1} \propto \frac{1}{T-\theta_3},\label{eq:scaling-cubic}
\end{eqnarray}
where $\theta_3~(>0)$ is proportional to $b$ in a mean-field argument for clean systems.~\cite{Kobler96}
This allows to estimate the spin-lattice coupling in experiments.
It is, however, unclear how the randomness $\Delta$ affects this high-$T$ behavior.

\begin{figure}[ht]
 \centering
 \includegraphics[width=0.4\textwidth,clip]{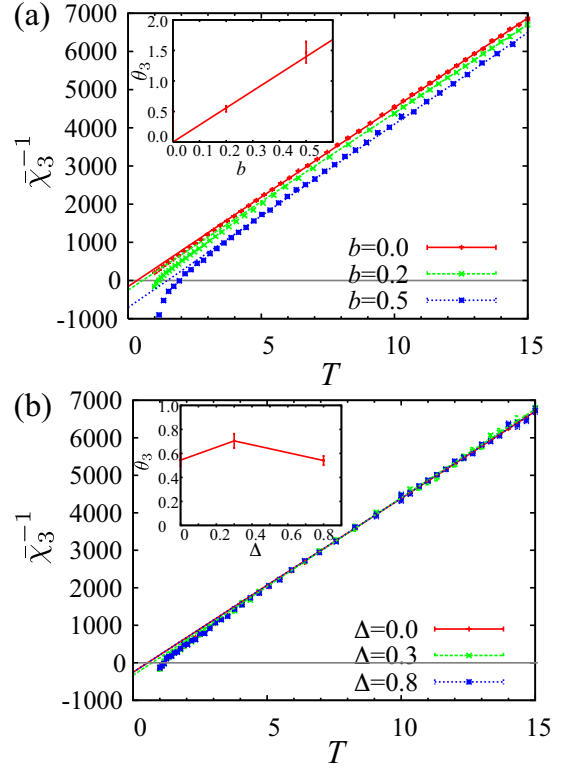}
 \caption{(Color online) 
 $T$ dependence of the inverse of the cubic susceptibility $\ncb$ calculated at $J_2^\mathrm{coop}=0$. The system size is $L=1$ (128 spins).
 (a) The data for different $b$ at $\Delta=0$.
 The straight lines denote the fits by high-$T$ asymptotic behavior in Eq.~(\ref{eq:scaling-cubic}).
 The inset shows the estimated $\theta_3$ is shown as a function of $b$.
 (b) The data for different $\Delta$ at $b=0.2$.
 The straight lines denote the fits by Eq.~(\ref{eq:scaling-cubic}).
 The inset shows the $\Delta$ dependence of the estimated $\theta_3$.
 }
 \label{fig:chi3-bar}
\end{figure}

Figure~\ref{fig:chi3-bar}(a) shows $\ncb$ calculated for different values of $b$ at $\Delta=0$.
As expected, the data obey the Curie-Weiss-like law at high $T$.
The estimated value of $\theta_3$ by the fitting by Eq.~(\ref{eq:scaling-cubic}) linearly increases with $b$, as plotted in the inset of Fig.~\ref{fig:chi3-bar}(a).
We further show the $T$ dependence of $\ncb$ while varying $\Delta$ at $b=0.2$ in Fig.~\ref{fig:chi3-bar}(b).
$\ncb$ is insensitive to $\Delta$. 
Indeed, as shown in the inset of Fig.~\ref{fig:chi3-bar}(b), the estimates of $\theta_3$ are almost independent on $\Delta$.
Our results indicate that the strength of the spin-lattice coupling $b$ can be measured by nonlinear susceptibility measurements even in the presence of randomness.

An alternative measure of the nonlinearity in the magnetic behavior is the nonlinear susceptibility
$\chi_3 (\equiv \partial^3 m/\partial H^3)$ defined in Eq.~(\ref{eq:def-chi3}).
While this quantity displays a positive divergence at a nematic transition,~\cite{Ramirez92} it shows a negative divergence at a canonical SG transition as~\cite{Fisher88,Chalupa77}
\begin{eqnarray}
  \nc &\propto& -|T-\TSG|^{-\gamma},
\end{eqnarray}
with a positive $\gamma$. 
A negative divergence of $\nc$ at $\TSG$ was also reported for one of geometrically frustrated SG materials, Y$_2$Mo$_2$O$_7$.~\cite{Gingras97}
Thus, it is of interest how $\chi_3$ behaves at the concomitant transition of nematic and SG in the plateau regime in our model.

\begin{figure}[ht]
 \centering
 \includegraphics[width=0.4\textwidth,clip]{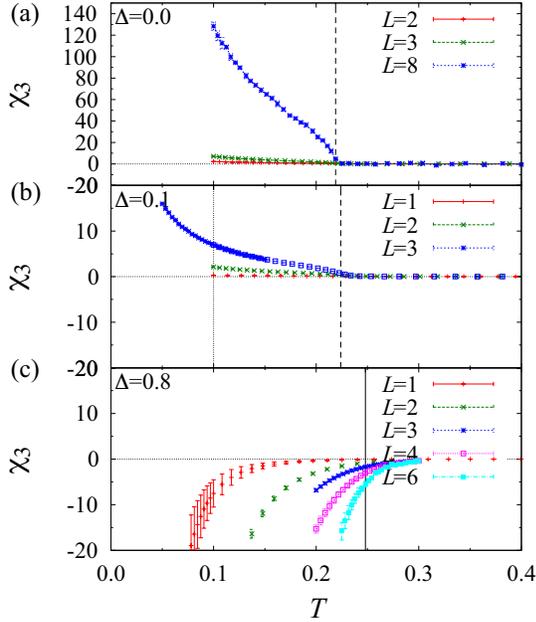}
 \caption{(Color online) $T$ dependence of the nonlinear susceptibility $\nc$ calculated for $b=0.2$ and $J_2^\mathrm{coop}=0$ at (a) $\Delta=0.0$, (b) $\Delta=0.1$, and (c) $\Delta=0.8$.
 The vertical dashed, dotted, solid lines denote the nematic transition temperature, the SG transition temperature, and the transition temperature of the concomitant transition, repectively (see Table~\ref{table:fss-linear}).
 }
 \label{fig:chi3}
\end{figure}

Figure~\ref{fig:chi3} shows $\nc$ calculated for different values of $\Delta$ at $b=0.2$.
The result in the absence of randomness is shown in Fig.~\ref{fig:chi3}(a). 
Below $\Tc$, $\nc$ increases as the system size increases, indicating that $\nc$ diverges to $+\infty$ at the nematic transition. 
Similar behavior is observed in the nematic transition in the linear regime, as shown in Fig.~\ref{fig:chi3}(b).
In contrast, $\nc$ shows a negative divergence at the concomitant transition in the plateau regime.
This is clearly seen in the result at $\Delta=0.8$ in Fig.~\ref{fig:chi3}(c).

\begin{figure}[ht]
 \centering
 \includegraphics[width=0.4\textwidth,clip]{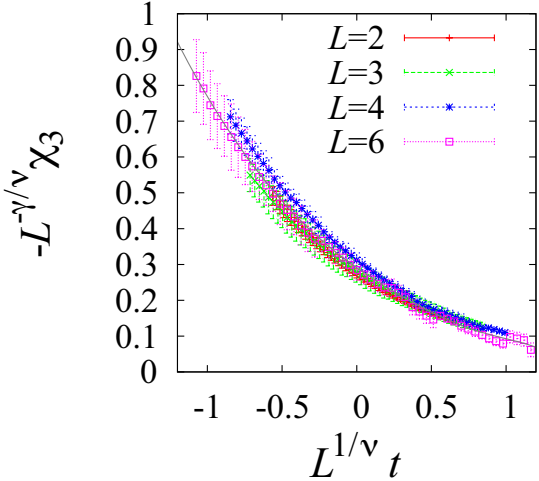}
 \caption{(Color online) Scaling collapse of the nonlinear  susceptibility $\nc$ at $\Delta=0.8$. The data are taken from Fig.~\ref{fig:chi3}(c). 
 We obtained the best fit with $\gamma=2.8(5) $ and $\nu=1.7(3)$.
 }
 \label{fig:chi3-collapse}
\end{figure}

To confirm the negative divergence of $\chi_3$ in the bulk limit, we perform a finite-size scaling analysis by assuming
\begin{eqnarray}
	\nc &=& -L^{\gamma/\nu} f(L^{1/\nu}t).
\end{eqnarray}
Here, we fixed $\TSG$ at the value obtained by the scaling analysis of $\chiSG$ (see Table~\ref{table:fss-linear}).
Figure~\ref{fig:chi3-collapse} shows the result at $\Delta=0.8$ for $L=2$--$6$.
The data collapse onto a single curve within error bars, indicating that $\chi_3$ continuously diverges at $\TSG$.
The large error bars of $\chi_3$ are because the MC sampling of the fourth-order moment in Eq.~(\ref{eq:MC-nc}) suffers from bad statistics.
The exponents are estimated to be $\gamma=2.8\pm 0.5$ and $\nu=1.7\pm 0.3$.
These results support that $\nc$ exhibits a negative divergence with $\gamma>0$ at the concomitant transition as in the case of the canonical SG.
This is consistent with the experimental result for Y$_2$Mo$_2$O$_7$ as we will discuss in Sec.~\ref{sec:comparison-with-exp}.

%% file: magnetic_field.tex
\section{Effects of magnetic field}\label{sec:mag-field}
In this section, we discuss effects of an external magnetic field on the concomitant transition in the plateau regime.
The canonical SG is sensitively affected by a magnetic field, even when the energy scale of the field is considerably smaller than $\TSG$ at $H$=0.
For instance, the transition temperature decreases rapidly for $H$ as ${\TSG(H)}/{\TSG(H=0)} -1 \propto -[{H}/{\TSG(H=0)}]^{2/3}$, which is called the Almeida-Thouless line, at the mean-field level.~\footnote{For more detail, please refer to Sec.~II.C.3 in Ref.~\onlinecite{Binder86}}
Although effects of a magnetic field on canonical SG beyond the mean-field approximation are still under investigation,
it was reported that a weak magnetic field destroys SG for an Ising three-dimensional Edwards-Anderson model.~\cite{Young04,Sasaki07}
In contrast, as mentioned in Sec.~I, the SG transition in Y$_2$Mo$_2$O$_7$ is less susceptible to an external magnetic field.~\cite{Silverstein13} 
Thus, it is of interest to clarify how the specific heat and the magnetic susceptibility behave in an external magnetic field for understanding SG behavior in frustrated SG magnets.  

\begin{figure}
 \centering
 \includegraphics[width=.4\textwidth,clip]{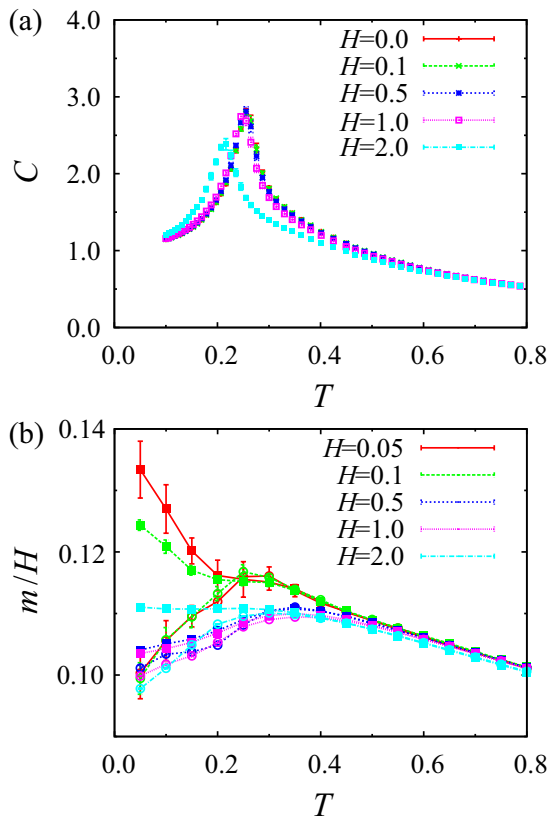}
 \caption{(Color online)
 (a) $H$ dependence of the specific heat $C$ at $\Delta=0.5$.
 (b) $H$ dependence of the ZFC (open circles) and FC (filled squares) susceptibilities at $\Delta=0.8$. See also Fig.~\ref{fig:chi-ZFC-FC}.
 The data are calculated at $\Jcoop=0$ for the system size $L=3$.
 }
 \label{fig:finite-H}
\end{figure}

Figure~\ref{fig:finite-H}(a) shows the specific heat calculated at different magnetic fields at $b=0.2$, $\Jcoop=0$, and $\Delta=0.5$.
At $H=0$, the $T$ dependence of the specific heat displays a peak around $\TSG$.
For $H>0$, the peak shows less change in its position and height up to $H \simeq 1$, whose energy scale is much larger than $\TSG\simeq 0.25$.
The peak is slightly broadened and shifted to a lower $T$ for $H \gtrsim 1$, as shown in Fig.~\ref{fig:finite-H}(a). 

Figure~\ref{fig:finite-H}(b) shows the $H$ dependence of the ZFC and FC susceptibilities calculated at $\Delta=0.8$.
The magnetic susceptibilities were calculated in the same procedures as in Sec.~\ref{sec:fc}.
At $H=0$, the ZFC susceptibility shows a cusp around $\TSG$,
below which the ZFC and FC susceptibilities split.
The temperature where the split takes place remains almost unchanged up to $H=2.0$, as shown in Fig.~\ref{fig:finite-H}(b). 
We note that the split becomes smaller as $H$ increases, but it increases for $H \gtrsim 1$, as shown in Fig.~\ref{fig:finite-H}(b). 

The results show that the concomitant transition, i.e., 
the peak in the specific heat and the hysteresis in the susceptibility, are robust against an applied magnetic field.
This is in clear contrast to the canonical SG which is strongly disturbed by the magnetic field.
The results well explain the robust SG behavior observed in Y$_2$Mo$_2$O$_7$.~\cite{Silverstein13}

%% file: spin_dynamics.tex
\section{Single-spin-flip dynamics in the nematic phase}\label{sec:spin-relax}
As shown in Sec.~\ref{sec:pd-J2-off}, the spin-lattice coupling induces the nematic phase in the weakly disordered region as well as in the clean case in the absence of the cooperative coupling $\Jcoop$.
The system exhibits the semidiscrete spin-ice macroscopic degeneracy in the nematic phase, which may lead to peculiar spin relaxation.

Indeed, spin dynamics characteristic to the spin-ice manifold has been extensively investigated 
for understanding magnetic properties in 4$f$ pyrochlores, such as Dy$_2$Ti$_2$O$_7$ and Ho$_2$Ti$_2$O$_7$ (refer to Ref.~\onlinecite{Gardner10} for a review). 
The Ising dipolar spin-ice model, which includes ferromagnetic nearest-neighbor interactions and long-range dipolar interactions as well as the local [111] easy-axis anisotropy,
is considered to be the relevant model for these compounds.
When the long-range dipolar interections are omitted, the low-$T$ state of the system suffers from the spin-ice macroscopic degeneracy.
That is, the system has the macroscopic number of degenerate ground states, which are separated by large energy barriers on the order of the exchange interaction.
Although the long-range parts of the dipolar interactions lift the spin-ice degeneracy, standard single-spin-flip MC simulations do not observe any transition down to low $T$.~\cite{Hertog00}
This is due to the freezing of the MC dynamics in the spin-ice manifold:
once the system enters into one of the spin-ice degenerate states, the system is dynamically trapped in the local minimum~\cite{Melko01,Melko04}.
On the other hand, the real materials, such as Dy$_2$Ti$_2$O$_7$ and Ho$_2$Ti$_2$O$_7$, do not show any magnetic transition down to the lowest $T$ in experiments.~\cite{Harris97,Ramirez99}
Furthermore, low-$T$ specific-heat measurements are in good agreement with results of the single-spin-flip MC simulations.~\cite{Ramirez99, Bramwell01}
These results indicate that they are in a nonequilibrium state and the spin dynamics becomes local at low $T$.
Indeed, nonequilibrium dynamics of local excitations from spin-ice states (monopoles) has been extensively studied
for understanding magnetic and thermodynamic properties in dipolar spin-ice materials.~\cite{Castelnovo10,Castelnovo11,Jaubert09}

For the present model with the biquadratic interaction, the spin collinearity emerges in the nematic phase in the weakly disordered region (see Fig.~\ref{fig:multivalley}). 
Since this enforces spins to follow the ice rule, similar dynamical freezing of spin dynamics is expected in the nematic phase.
To see how spin dynamics freezes as $T$ is lowered, we perform MC simulation only with the single-spin-flip update.
Spin relaxation is measured by the autocorrelation function in the form
\begin{eqnarray}
 A(n) &=&
 C_\mathrm{norm}
 \Big\{
 \Big\langle \big(\sum_i \vec{S}_i (n_0) \cdot \vec{S}_i (n_0 + n)\big)^2 \Big\rangle_\Delta-\nonumber\\
 &&\Big\langle \big(\sum_i \vec{S}_i (n_0) \cdot \vec{S}_i (n_0 + \infty)\big)^2 \Big\rangle_\Delta \Big\},
 \label{eq:autocorr}
\end{eqnarray}
where $\vec{S}_i(n)$ is the spin at $i$th site in the sample at $n$th MC step.
We take $A(0)=1$ ($C_\mathrm{norm}$ is a normalization factor).
The autocorrelation fucntion measures the correlatiton between the MC samples in the interval $n$.
We calculate this quantity after the Monte Carlo dynamics is thermalized at each $T$.
In this study, we fix $n_0$ to the first MC step after the thermalization.

\begin{figure}[h]
 \centering
 \includegraphics[width=.4\textwidth,clip]{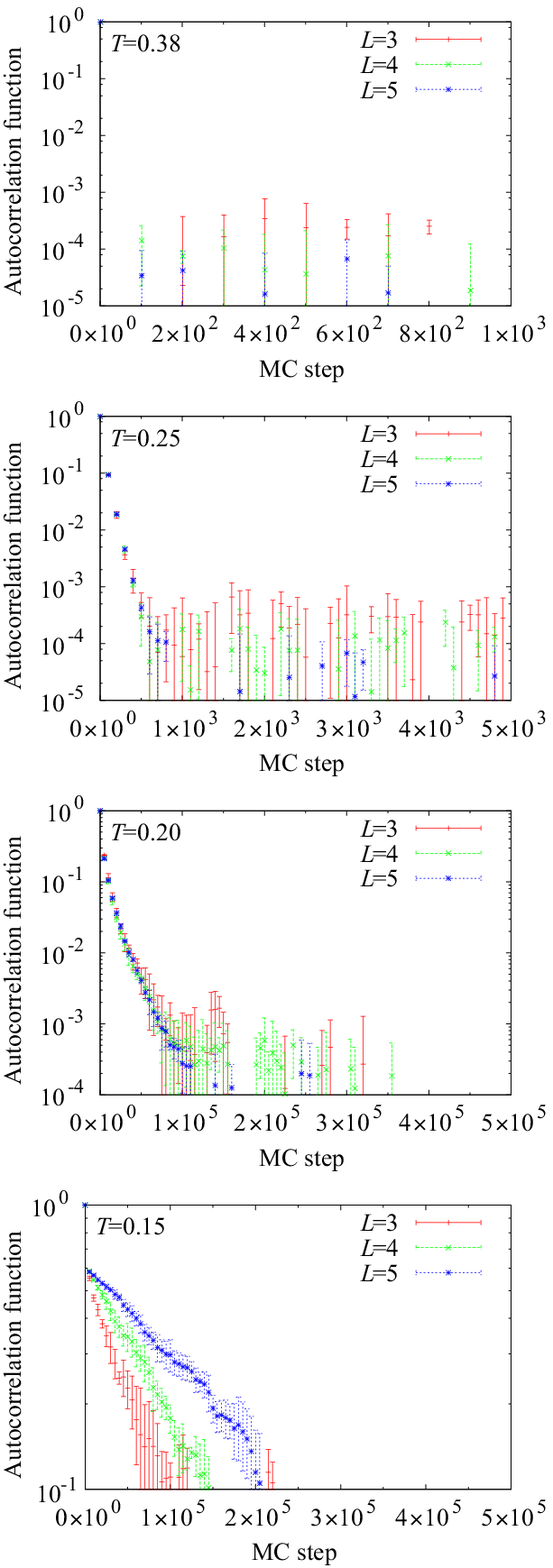}
 \caption{(Color online) Autocorrelation functions calculated with $\Delta=0.1$, $b=0.2$, and $J_2^\mathrm{coop}=0.0$ (linear regime).
 The MC dynamics exhibits a dynamical freezing below $\Tc\simeq b$.
 }
 \label{fig:relax-D0.1}
\end{figure}

Figure~\ref{fig:relax-D0.1} shows the autocorrelation functions calculated with $\Delta=0.1$.
We obtained essentially the same data for $\Delta=0$ (not shown).
At high $T>b$, e.g., $T=0.38$, the autocorrelation functions decay rapidly.
The nonzero asymptotic values $A(\infty)~(>0)$ are due to a finite-size effect, which vanish as $L$ increases.
At lower $T<\Tc\simeq 0.225$, the emergent multivalley structure is expected to prevent the single-spin-flip dynamics from exploring the whole manifold.
Indeed, the autocorrelation functions exhibit a severe freezing when entering the nematic phase;
the autocorrelation functions do not vanish even after $4\times 10^5$ MC steps at $T=0.15$.
Note that this $T$ range is still higher than $\TSG\simeq 0.102$ at $\Delta=0.1$.
These results indicate that the spin freezing may appear at $\Tc\simeq b$ even for negligibly small randomness when only single-spin-flip dynamics is considered.
In the next section, we discuss implications of these results in understanding of the robust SG behavior experimentally observed in frustrated magnets.

%% file: comparison.tex
\section{Comparisons with experiments}\label{sec:comparison-with-exp}
In this section, we discuss the results of the peculiar SG behavior induced by the spin-lattice coupling in comparison with experiments. 
Experimentally, even high-quality samples of the stoichiometric compound Y$_2$Mo$_2$O$_7$ show a SG transition. 
The robust SG behavior was recently observed also for a single crystal.~\cite{Silverstein13} 
A chemical disorder, introduced by, e.g., La substitution of Y, does not affect the critical temperature $\TSG$, while it significantly increases the Curie-Weiss temperature.~\cite{Sato87} 
Our results presented in this paper provide a way of understanding the peculiar SG behavior.
An important observation is that many experiments suggest a substantial bond disorder even in the stoichiometric samples without chemical disorder.~\cite{Booth00,Keren01,Sagi05,Greedan09}
The relevance of the spin-lattice coupling was also pointed out.~\cite{Ofer10}
Suppose that the compounds inevitably include a substantial disorder and are already in the plateau regime, 
they undergo a concomitant phase transition, and the critical temperature $\TSG$ can be large and remain almost constant against additional disorder, as discussed in Sec.~\ref{sec:plateau}.
In contrast, the Curie-Weiss temperature $\theta_\mathrm{CW}$ estimated above $\TSG$ changes depending on the additional disorder, as shown in Sec.~\ref{sec:mag-chi1}. 
In Sec.~\ref{sec:spin-relax}, we further showed that single-spin-flip dynamics freezes even in the weakly disordered regime once the system enters the nematic phase. 
Although it is not obvious how this slowing down is observed in experiments, 
the results suggest that, in the experimental time scale, the freezing SG behavior might be observed at around $T_c$, which is set by the spin-lattice coupling $b$, even if randomness is negligibly small.

Our results are also consistent with the experimental results for the magnetic specific heat. 
For Y$_2$Mo$_2$O$_7$, a broad peak was observed around $\TSG$ in the $T$ dependence of the specific heat.~\cite{Raju92} 
Furthermore, the broad peak was recently reported to be insensitive to an applied magnetic field.~\cite{Silverstein13}
These are in clear contrast to the canonical SG; the specific heat exhibits a cusp at a slightly higher temperature than $\TSG$, and $\TSG$ is sensitively suppressed by a magnetic field.~\cite{Binder86}
The peculiar behavior, however, is reproduced in our results including the effect of the spin-lattice coupling, as shown in Secs.~\ref{sec:plateau} and \ref{sec:mag-field}.
Further experiments on other SG materials and high-field measurements are desirable to clarify the nature of the SG.

Our study revealed that the concomitant transition is consistent with a second-order transition and is accompanied by the divergent behavior of $\nc$, i.e., $\nc\rightarrow -\infty$ as $T \to \TSG$. 
This behavior is consistent with the experimental result for Y$_2$Mo$_2$O$_7$; the SG transition is continuous and accompanied by the power-law divergence of $\nc \propto - (T-\TSG)^{-\gamma}$ with $\gamma \simeq 2.8$.~\cite{Gingras97}
The value of the critical exponent does not contradict with our estimate of $\gamma = 2.8 \pm 0.5$ obtained by the finite-size scaling of $\nc$ (see Fig.~\ref{fig:chi3-collapse}).

On the other hand, our results with the cooperative coupling of local lattice distortions, i.e., for $J_2^\mathrm{coop}\neq 0$, qualitatively
explain the phase competition between the spin-lattice ordered phase and SG phase in the case of (Zn$_{1-x}$Cd$_x$)Cr$_2$O$_4$. 
In these compounds, the doping of Cd quickly destroys the N\'{e}el order with uniform lattice distortions at $x \simeq 0.03$, and induces SG behavior at $\TSG \simeq 10$~K; the value of $\TSG$ remains unchanged up to $x \sim 0.1$.
Similar phase competition and robust behavior of $\TSG$ are also seen in the spinel CoAl$_{2}$O$_{4}$.
In CoAl$_{2}$O$_{4}$, the magnetic phase diagram is controlled by intersite mixing between magnetic Co and nonmagnetic Al sites, $\eta$, as (Co$_{1-\eta}$Al$_\eta$)[Al$_{2-\eta}$Co$_\eta$]O$_4$.~\cite{Hanashima13}
For $\eta \gtrsim 0.08$, the system shows a SG transition at $\TSG\simeq 4.5$ K,
which is almost constant for $\eta \le 0.153$.
Recently, it was reported that a high-quality sample with $\eta=0.057(20)$ shows a N\'{e}el transition at $\Tc=9.8(2)$ K.~\cite{Roy13}
Thus, in both cases of (Zn$_{1-x}$Cd$_x$)Cr$_2$O$_4$ and (Co$_{1-\eta}$Al$_\eta$)[Al$_{2-\eta}$Co$_\eta$]O$_4$, the experimental phase diagrams are consistent with our results in Sec.~\ref{sec:pd-J2-on}.
Further experiments on the magnetic susceptibility and specific heat are desired to clarify the nature of the SG transition and the role of the spin-lattice coupling.

%% file: summary.tex
\section{Summary and conclusion remarks}
In this paper, we have investigated effects of the spin-lattice coupling on SG transitions in bond-disordered Heisenberg pyrochlore antiferromagnets coupled with local lattice distortions by Monte Carlo simulations.
The coupling to lattice distortions is taken into account in the effective spin-only models in the form of the nearest-neighbor biquadratic interaction and further-neighbor bilinear interactions. 
The latter originates in the cooperative aspect of the local lattice distortions.

Let us first summarize our findings for the case with the nearest-neighbor couplings only.
The disorder($\Delta$)--temperature($T$) phase diagram exhibits the following characteristics: 
In the weakly disordered regime, the SG transition temperature $\TSG$ grows linearly with $\Delta$,
showing a remarkable enhancement by the coupling to local lattice distortions $b$
As $\Delta$ increases, the system enters the plateau regime where the concomitant transition of SG and nematic order takes place at $\TSG \simeq b$, being almost independent of $\Delta$. 
We have also found that the Curie-Weiss temperature estimated above $\TSG$ sensitively changes as a function of $\Delta$.
All these results well explain the peculiar SG behavior observed in $R_2$Mo$_2$O$_7$. 

We have further investigated thermodynamic properties near the concomitant transition.
We found that the concomitant transition has the following aspects that resemble the canonical SG behavior:
the nonlinear susceptibility $\chi_3$ displays a negative divergence at the concomitant transition,
and the magnetic susceptibility shows hysteresis behavior between the FC and ZFC measurements below $\TSG$.
On the other hand, the concomitant transition has the following unconventional characteristics: 
the specific heat $C$ displays a broad peak around $\TSG$, 
and the transition is robust against an external magnetic field. 
These results are also consistent with the experimental observations for Y$_2$Mo$_2$O$_7$.~\cite{Raju92,Gingras97,Silverstein13}
High magnetic field measurements are desirable to further understand the SG behavior.

Furthermore, we have investigated effects of the spin-lattice coupling on the nonlinearlity of the magnetic susceptibility in the high-temperature paramagnetic phase.
We have shown that the high-$T$ measurement of the cubic susceptibility gives a good measure of the strength of the spin-lattice coupling $b$ even in the presence of disorder.

We have also studied spin relaxation in the nematic phase in the weakly disordered regime.
We have shown that single-spin-flip dynamics freezes once the system enters the nematic phase even if the randomness is negligibly small.
This may explain the SG behavior experimentally observed in high-quality samples of many frustrated magnets.

In the case with the cooperative coupling between local lattice distortions, as discussed in the previous paper,~\cite{Shinaoka10b}
the cooperative coupling $\Jcoop$ results in the phase competition between the spin-lattice phase and the SG phase.
We have presented that the critical properties as well as the behavior of $\TSG$ are similar to the case with $\Jcoop=0$.
The results give a reasonable explanation for the phase competition observed in Zn spinels.

Finally, let us discuss future directions of the study of the SG behavior in frustrated magnets.
Y$_2$Mo$_2$O$_7$ and Lu$_2$Mo$_2$O$_7$ show peculiar $T^2$-temperature dependence in the specific heat at low temperatures below the spin-glass transition temperature.~\cite{Silverstein13,Clark14}
This is in clear contrast to the canonical SG in which the specific heat shows linear temperature dependence.
It was speculated that the orbital degree of freedom plays an important role in this unusual behavior.~\cite{Silverstein13}
Similar $T^2$-temperature dependence in the specific heat, however, was observed for some cubic spinels CoAl$_2$O$_4$ and FeAl$_2$O$_4$
with no orbital degree of freedom.~\cite{Tristan05}
It is left for future study to clarify the effects of the coupling between spin, orbital, and lattice on the low-temperature behavior in the specific heat.

Recent first-principles studies indicate the substantial role of the orbital degree freedom in the magnetism for  Y$_2$Mo$_2$O$_7$.~\cite{Shinaoka-YMO,Silverstein13}
In particular, two of the authors and co-workers showed that the effective spin interactions are strongly anisotropic in spin space due to the strong coupling between spin and orbital through the relativistic spin-orbit coupling.~\cite{Shinaoka-YMO}
It is of great interest to investigate how such magnetic anisotropy affects the scenario in the present study. 

\begin{acknowledgments}
We thank T. Kato, H. Kawamura, K. Penc, N. Shannon, and H. J. Silverstein for fruitful discussion.
Numerical calculation was partly carried out at the Supercomputer Center, ISSP, Univ. of Tokyo. 
This work was supported by the Strategic Programs for Innovative Research (SPIRE), MEXT, and the Computational Materials Science Initiative (CMSI), Japan.
\end{acknowledgments}

\bibliography{ref}

%% file: appendix.tex
\section{Demonstration of the extended loop algorithm}\label{sec:lm-demo}
\begin{figure}[ht]
 \centering
 \resizebox{0.4\textwidth}{!}{\includegraphics{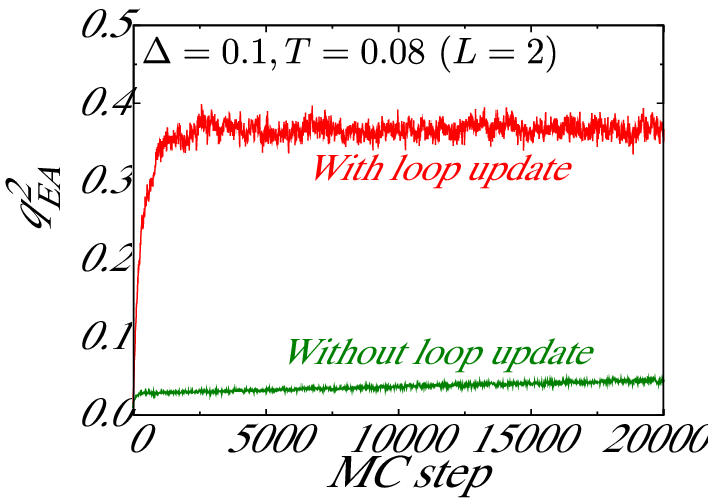}}
 \caption{Comparison of thermalization processes of $\qEA$ 
 from a disordered configuration with and without using the extended loop update.
 The data are taken at $\Delta=0.1$ and $T=0.08$ for the model (\ref{eq:Ham2}) with $b=0.2$ and $J_2^\mathrm{coop}=0$ in the system size $L=2$.
 The relaxation is remarkably accelerated by the loop update.}
 \label{fig:lm-demo}
\end{figure}

In Fig.~\ref{fig:lm-demo}, 
we compare thermalization processes of the Edwards-Anderson order parameter $\qEA$ in Eq.~(\ref{eq:qEA}) with and without the loop update at $\Delta=0.1$ and $T=0.08$ slightly below $\TSG$ [see Fig.~\ref{fig:pd}(a)].
We take 16 temperature points uniformly distributed in the range of $0.08\le T\le 0.2$ for the exchange MC method.
The system size is $L=2$, i.e., $\Ns=128$ spins.
As shown in Fig.~\ref{fig:lm-demo}, the MC dynamics without the loop update suffers from severe slowing down;
it is extremely hard to thermalize the single-spin-flip MC dynamics despite the small system size.
When the loop flip is turned on,
the thermalization process is greatly accelerated.
The MC dynamics quickly reaches thermal equilibrium within $2.5\times 10^3$ MC steps as demonstrated in Fig.~\ref{fig:lm-demo}.
The results clearly show the advantage of the extended loop algorithm in investigating the low-$T$ properties of the present model.

%% file: SG-fullpaper.bbl
\begin{thebibliography}{86}
\expandafter\ifx\csname natexlab\endcsname\relax\def\natexlab#1{#1}\fi
\expandafter\ifx\csname bibnamefont\endcsname\relax
  \def\bibnamefont#1{#1}\fi
\expandafter\ifx\csname bibfnamefont\endcsname\relax
  \def\bibfnamefont#1{#1}\fi
\expandafter\ifx\csname citenamefont\endcsname\relax
  \def\citenamefont#1{#1}\fi
\expandafter\ifx\csname url\endcsname\relax
  \def\url#1{\texttt{#1}}\fi
\expandafter\ifx\csname urlprefix\endcsname\relax\def\urlprefix{URL }\fi
\providecommand{\bibinfo}[2]{#2}
\providecommand{\eprint}[2][]{\url{#2}}

\bibitem[{\citenamefont{Ramirez}(1994)}]{Ramirez94}
\bibinfo{author}{\bibfnamefont{A.~P.} \bibnamefont{Ramirez}},
  \bibinfo{journal}{Annual Review of Materials Science}
  \textbf{\bibinfo{volume}{24}}, \bibinfo{pages}{453} (\bibinfo{year}{1994}),
  \eprint{http://www.annualreviews.org/doi/pdf/10.1146/annurev.ms.24.080194.00%
2321},
  \urlprefix\url{http://www.annualreviews.org/doi/abs/10.1146/annurev.ms.24.08%
0194.002321}.

\bibitem[{\citenamefont{Binder and Young}(1986)}]{Binder86}
\bibinfo{author}{\bibfnamefont{K.}~\bibnamefont{Binder}} \bibnamefont{and}
  \bibinfo{author}{\bibfnamefont{A.~P.} \bibnamefont{Young}},
  \bibinfo{journal}{Rev. Mod. Phys.} \textbf{\bibinfo{volume}{58}},
  \bibinfo{pages}{801} (\bibinfo{year}{1986}).

\bibitem[{\citenamefont{Ruderman and Kittel}(1954)}]{Ruderman54}
\bibinfo{author}{\bibfnamefont{M.~A.} \bibnamefont{Ruderman}} \bibnamefont{and}
  \bibinfo{author}{\bibfnamefont{C.}~\bibnamefont{Kittel}},
  \bibinfo{journal}{Phys. Rev.} \textbf{\bibinfo{volume}{96}},
  \bibinfo{pages}{99} (\bibinfo{year}{1954}),
  \urlprefix\url{http://link.aps.org/doi/10.1103/PhysRev.96.99}.

\bibitem[{\citenamefont{Kasuya}(1956)}]{Kasuya56}
\bibinfo{author}{\bibfnamefont{T.}~\bibnamefont{Kasuya}},
  \bibinfo{journal}{Progress of Theoretical Physics}
  \textbf{\bibinfo{volume}{16}}, \bibinfo{pages}{45} (\bibinfo{year}{1956}),
  \urlprefix\url{http://ptp.ipap.jp/link?PTP/16/45/}.

\bibitem[{\citenamefont{Yosida}(1957{\natexlab{a}})}]{Yoshida57a}
\bibinfo{author}{\bibfnamefont{K.}~\bibnamefont{Yosida}},
  \bibinfo{journal}{Phys. Rev.} \textbf{\bibinfo{volume}{106}},
  \bibinfo{pages}{893} (\bibinfo{year}{1957}{\natexlab{a}}),
  \urlprefix\url{http://link.aps.org/doi/10.1103/PhysRev.106.893}.

\bibitem[{\citenamefont{Yosida}(1957{\natexlab{b}})}]{Yoshida57b}
\bibinfo{author}{\bibfnamefont{K.}~\bibnamefont{Yosida}},
  \bibinfo{journal}{Phys. Rev.} \textbf{\bibinfo{volume}{107}},
  \bibinfo{pages}{396} (\bibinfo{year}{1957}{\natexlab{b}}),
  \urlprefix\url{http://link.aps.org/doi/10.1103/PhysRev.107.396}.

\bibitem[{\citenamefont{Wannier}(1950)}]{Wannier50}
\bibinfo{author}{\bibfnamefont{G.~H.} \bibnamefont{Wannier}},
  \bibinfo{journal}{Phys. Rev.} \textbf{\bibinfo{volume}{79}},
  \bibinfo{pages}{357} (\bibinfo{year}{1950}),
  \urlprefix\url{http://link.aps.org/doi/10.1103/PhysRev.79.357}.

\bibitem[{\citenamefont{Wannier}(1973)}]{Wannier73}
\bibinfo{author}{\bibfnamefont{G.~H.} \bibnamefont{Wannier}},
  \bibinfo{journal}{Phys. Rev. B} \textbf{\bibinfo{volume}{7}},
  \bibinfo{pages}{5017} (\bibinfo{year}{1973}),
  \urlprefix\url{http://link.aps.org/doi/10.1103/PhysRevB.7.5017}.

\bibitem[{\citenamefont{Houtappel}(1950)}]{Houtappel50}
\bibinfo{author}{\bibfnamefont{R.}~\bibnamefont{Houtappel}},
  \bibinfo{journal}{Physica} \textbf{\bibinfo{volume}{16}}, \bibinfo{pages}{425
  } (\bibinfo{year}{1950}), ISSN \bibinfo{issn}{0031-8914},
  \urlprefix\url{http://www.sciencedirect.com/science/article/pii/003189145090%
1303}.

\bibitem[{\citenamefont{Husimi and Sy{\^o}zi}(1950)}]{Husimi50}
\bibinfo{author}{\bibfnamefont{K.}~\bibnamefont{Husimi}} \bibnamefont{and}
  \bibinfo{author}{\bibfnamefont{I.}~\bibnamefont{Sy{\^o}zi}},
  \bibinfo{journal}{Progress of Theoretical Physics}
  \textbf{\bibinfo{volume}{5}}, \bibinfo{pages}{177} (\bibinfo{year}{1950}),
  \eprint{http://ptp.oxfordjournals.org/content/5/2/177.full.pdf+html},
  \urlprefix\url{http://ptp.oxfordjournals.org/content/5/2/177.abstract}.

\bibitem[{\citenamefont{Diep}(2005)}]{Diep05}
\bibinfo{author}{\bibfnamefont{H.}~\bibnamefont{Diep}},
  \emph{\bibinfo{title}{Frustrated Spin Systems}} (\bibinfo{publisher}{World
  Scientific}, \bibinfo{address}{Singapore}, \bibinfo{year}{2005}).

\bibitem[{\citenamefont{Ramirez et~al.}(1990)\citenamefont{Ramirez, Espinosa,
  and Cooper}}]{Ramirez90}
\bibinfo{author}{\bibfnamefont{A.~P.} \bibnamefont{Ramirez}},
  \bibinfo{author}{\bibfnamefont{G.~P.} \bibnamefont{Espinosa}},
  \bibnamefont{and} \bibinfo{author}{\bibfnamefont{A.~S.}
  \bibnamefont{Cooper}}, \bibinfo{journal}{Phys. Rev. Lett.}
  \textbf{\bibinfo{volume}{64}}, \bibinfo{pages}{2070} (\bibinfo{year}{1990}).

\bibitem[{\citenamefont{Martinho et~al.}(2001)\citenamefont{Martinho, Moreno,
  Sanjurjo, Rettori, Garc\'\i{}a-Adeva, Huber, Oseroff, Ratcliff, Cheong,
  Pagliuso et~al.}}]{Martinho01}
\bibinfo{author}{\bibfnamefont{H.}~\bibnamefont{Martinho}},
  \bibinfo{author}{\bibfnamefont{N.~O.} \bibnamefont{Moreno}},
  \bibinfo{author}{\bibfnamefont{J.~A.} \bibnamefont{Sanjurjo}},
  \bibinfo{author}{\bibfnamefont{C.}~\bibnamefont{Rettori}},
  \bibinfo{author}{\bibfnamefont{A.~J.} \bibnamefont{Garc\'\i{}a-Adeva}},
  \bibinfo{author}{\bibfnamefont{D.~L.} \bibnamefont{Huber}},
  \bibinfo{author}{\bibfnamefont{S.~B.} \bibnamefont{Oseroff}},
  \bibinfo{author}{\bibfnamefont{W.}~\bibnamefont{Ratcliff}},
  \bibinfo{author}{\bibfnamefont{S.-W.} \bibnamefont{Cheong}},
  \bibinfo{author}{\bibfnamefont{P.~G.} \bibnamefont{Pagliuso}},
  \bibnamefont{et~al.}, \bibinfo{journal}{Phys. Rev. B}
  \textbf{\bibinfo{volume}{64}}, \bibinfo{pages}{024408}
  (\bibinfo{year}{2001}).

\bibitem[{\citenamefont{Tristan et~al.}(2005)\citenamefont{Tristan, Hemberger,
  Krimmel, Krug~von Nidda, Tsurkan, and Loidl}}]{Tristan05}
\bibinfo{author}{\bibfnamefont{N.}~\bibnamefont{Tristan}},
  \bibinfo{author}{\bibfnamefont{J.}~\bibnamefont{Hemberger}},
  \bibinfo{author}{\bibfnamefont{A.}~\bibnamefont{Krimmel}},
  \bibinfo{author}{\bibfnamefont{H.-A.} \bibnamefont{Krug~von Nidda}},
  \bibinfo{author}{\bibfnamefont{V.}~\bibnamefont{Tsurkan}}, \bibnamefont{and}
  \bibinfo{author}{\bibfnamefont{A.}~\bibnamefont{Loidl}},
  \bibinfo{journal}{Phys. Rev. B} \textbf{\bibinfo{volume}{72}},
  \bibinfo{pages}{174404} (\bibinfo{year}{2005}).

\bibitem[{\citenamefont{Gardner et~al.}(2010)\citenamefont{Gardner, Gingras,
  and Greedan}}]{Gardner10}
\bibinfo{author}{\bibfnamefont{J.~S.} \bibnamefont{Gardner}},
  \bibinfo{author}{\bibfnamefont{M.~J.~P.} \bibnamefont{Gingras}},
  \bibnamefont{and} \bibinfo{author}{\bibfnamefont{J.~E.}
  \bibnamefont{Greedan}}, \bibinfo{journal}{Rev. Mod. Phys.}
  \textbf{\bibinfo{volume}{82}}, \bibinfo{pages}{53} (\bibinfo{year}{2010}).

\bibitem[{\citenamefont{Munenaka and Sato}(2006)}]{Munekata06}
\bibinfo{author}{\bibfnamefont{T.}~\bibnamefont{Munenaka}} \bibnamefont{and}
  \bibinfo{author}{\bibfnamefont{H.}~\bibnamefont{Sato}},
  \bibinfo{journal}{Journal of the Physical Society of Japan}
  \textbf{\bibinfo{volume}{75}}, \bibinfo{pages}{103801}
  (\bibinfo{year}{2006}).

\bibitem[{\citenamefont{Zhou et~al.}(2008)\citenamefont{Zhou, Wiebe, Harter,
  Dalal, and Gardner}}]{Zhou08}
\bibinfo{author}{\bibfnamefont{H.~D.} \bibnamefont{Zhou}},
  \bibinfo{author}{\bibfnamefont{C.~R.} \bibnamefont{Wiebe}},
  \bibinfo{author}{\bibfnamefont{A.}~\bibnamefont{Harter}},
  \bibinfo{author}{\bibfnamefont{N.~S.} \bibnamefont{Dalal}}, \bibnamefont{and}
  \bibinfo{author}{\bibfnamefont{J.~S.} \bibnamefont{Gardner}},
  \bibinfo{journal}{Journal of Physics: Condensed Matter}
  \textbf{\bibinfo{volume}{20}}, \bibinfo{pages}{325201}
  (\bibinfo{year}{2008}).

\bibitem[{\citenamefont{Reimers}(1992)}]{Reimers92}
\bibinfo{author}{\bibfnamefont{J.~N.} \bibnamefont{Reimers}},
  \bibinfo{journal}{Phys. Rev. B} \textbf{\bibinfo{volume}{45}},
  \bibinfo{pages}{7287} (\bibinfo{year}{1992}).

\bibitem[{\citenamefont{Moessner and
  Chalker}(1998{\natexlab{a}})}]{Moessner98a}
\bibinfo{author}{\bibfnamefont{R.}~\bibnamefont{Moessner}} \bibnamefont{and}
  \bibinfo{author}{\bibfnamefont{J.~T.} \bibnamefont{Chalker}},
  \bibinfo{journal}{Phys. Rev. Lett} \textbf{\bibinfo{volume}{80}},
  \bibinfo{pages}{2929} (\bibinfo{year}{1998}{\natexlab{a}}).

\bibitem[{\citenamefont{Saunders and Chalker}(2007)}]{Saunders07}
\bibinfo{author}{\bibfnamefont{T.~E.} \bibnamefont{Saunders}} \bibnamefont{and}
  \bibinfo{author}{\bibfnamefont{J.~T.} \bibnamefont{Chalker}},
  \bibinfo{journal}{Phys. Rev. Lett.} \textbf{\bibinfo{volume}{98}},
  \bibinfo{pages}{157201} (\bibinfo{year}{2007}).

\bibitem[{\citenamefont{Andreanov et~al.}(2010)\citenamefont{Andreanov,
  Chalker, Saunders, and Sherrington}}]{Andreanov10}
\bibinfo{author}{\bibfnamefont{A.}~\bibnamefont{Andreanov}},
  \bibinfo{author}{\bibfnamefont{J.~T.} \bibnamefont{Chalker}},
  \bibinfo{author}{\bibfnamefont{T.~E.} \bibnamefont{Saunders}},
  \bibnamefont{and}
  \bibinfo{author}{\bibfnamefont{D.}~\bibnamefont{Sherrington}},
  \bibinfo{journal}{Phys. Rev. B} \textbf{\bibinfo{volume}{81}},
  \bibinfo{pages}{014406} (\bibinfo{year}{2010}).

\bibitem[{\citenamefont{Bellier-Castella
  et~al.}(2001)\citenamefont{Bellier-Castella, Gingras, Holdsworth, and
  R.}}]{Bellier-Castella01}
\bibinfo{author}{\bibfnamefont{L.}~\bibnamefont{Bellier-Castella}},
  \bibinfo{author}{\bibfnamefont{M.~J.~P.} \bibnamefont{Gingras}},
  \bibinfo{author}{\bibfnamefont{P.~C.~W.} \bibnamefont{Holdsworth}},
  \bibnamefont{and} \bibinfo{author}{\bibfnamefont{M.}~\bibnamefont{R.}},
  \bibinfo{journal}{Can. J. Phys.} \textbf{\bibinfo{volume}{79}},
  \bibinfo{pages}{1365} (\bibinfo{year}{2001}).

\bibitem[{\citenamefont{Gingras et~al.}(1997)\citenamefont{Gingras, Stager,
  Raju, Gaulin, and Greedan}}]{Gingras97}
\bibinfo{author}{\bibfnamefont{M.~J.~P.} \bibnamefont{Gingras}},
  \bibinfo{author}{\bibfnamefont{C.~V.} \bibnamefont{Stager}},
  \bibinfo{author}{\bibfnamefont{N.~P.} \bibnamefont{Raju}},
  \bibinfo{author}{\bibfnamefont{B.~D.} \bibnamefont{Gaulin}},
  \bibnamefont{and} \bibinfo{author}{\bibfnamefont{J.~E.}
  \bibnamefont{Greedan}}, \bibinfo{journal}{Phys. Rev. Lett.}
  \textbf{\bibinfo{volume}{78}}, \bibinfo{pages}{947} (\bibinfo{year}{1997}).

\bibitem[{\citenamefont{Greedan et~al.}(1986)\citenamefont{Greedan, Sato, Yan,
  and Razavi}}]{Greedan86}
\bibinfo{author}{\bibfnamefont{J.}~\bibnamefont{Greedan}},
  \bibinfo{author}{\bibfnamefont{M.}~\bibnamefont{Sato}},
  \bibinfo{author}{\bibfnamefont{X.}~\bibnamefont{Yan}}, \bibnamefont{and}
  \bibinfo{author}{\bibfnamefont{F.}~\bibnamefont{Razavi}},
  \bibinfo{journal}{Solid State Communications} \textbf{\bibinfo{volume}{59}},
  \bibinfo{pages}{895 } (\bibinfo{year}{1986}), ISSN \bibinfo{issn}{0038-1098}.

\bibitem[{\citenamefont{Gingras et~al.}(1996)\citenamefont{Gingras, Stager,
  Gaulin, Raju, and Greedan}}]{Gingras96}
\bibinfo{author}{\bibfnamefont{M.~J.~P.} \bibnamefont{Gingras}},
  \bibinfo{author}{\bibfnamefont{C.~V.} \bibnamefont{Stager}},
  \bibinfo{author}{\bibfnamefont{B.~D.} \bibnamefont{Gaulin}},
  \bibinfo{author}{\bibfnamefont{N.~P.} \bibnamefont{Raju}}, \bibnamefont{and}
  \bibinfo{author}{\bibfnamefont{J.~E.} \bibnamefont{Greedan}},
  \bibinfo{journal}{J. Appl. Phys. 79, 6170 (1996);}
  \textbf{\bibinfo{volume}{79}}, \bibinfo{pages}{6170} (\bibinfo{year}{1996}).

\bibitem[{\citenamefont{Silverstein et~al.}(2014)\citenamefont{Silverstein,
  Fritsch, Flicker, Hallas, Gardner, Qiu, Ehlers, Savici, Yamani, Ross
  et~al.}}]{Silverstein13}
\bibinfo{author}{\bibfnamefont{H.~J.} \bibnamefont{Silverstein}},
  \bibinfo{author}{\bibfnamefont{K.}~\bibnamefont{Fritsch}},
  \bibinfo{author}{\bibfnamefont{F.}~\bibnamefont{Flicker}},
  \bibinfo{author}{\bibfnamefont{A.~M.} \bibnamefont{Hallas}},
  \bibinfo{author}{\bibfnamefont{J.~S.} \bibnamefont{Gardner}},
  \bibinfo{author}{\bibfnamefont{Y.}~\bibnamefont{Qiu}},
  \bibinfo{author}{\bibfnamefont{G.}~\bibnamefont{Ehlers}},
  \bibinfo{author}{\bibfnamefont{A.~T.} \bibnamefont{Savici}},
  \bibinfo{author}{\bibfnamefont{Z.}~\bibnamefont{Yamani}},
  \bibinfo{author}{\bibfnamefont{K.~A.} \bibnamefont{Ross}},
  \bibnamefont{et~al.}, \bibinfo{journal}{Phys. Rev. B}
  \textbf{\bibinfo{volume}{89}}, \bibinfo{pages}{054433}
  (\bibinfo{year}{2014}),
  \urlprefix\url{http://link.aps.org/doi/10.1103/PhysRevB.89.054433}.

\bibitem[{\citenamefont{Clark et~al.}(2014)\citenamefont{Clark, Nilsen,
  Kermarrec, Ehlers, Knight, Harrison, Attfield, and Gaulin}}]{Clark14}
\bibinfo{author}{\bibfnamefont{L.}~\bibnamefont{Clark}},
  \bibinfo{author}{\bibfnamefont{G.~J.} \bibnamefont{Nilsen}},
  \bibinfo{author}{\bibfnamefont{E.}~\bibnamefont{Kermarrec}},
  \bibinfo{author}{\bibfnamefont{G.}~\bibnamefont{Ehlers}},
  \bibinfo{author}{\bibfnamefont{K.~S.} \bibnamefont{Knight}},
  \bibinfo{author}{\bibfnamefont{A.}~\bibnamefont{Harrison}},
  \bibinfo{author}{\bibfnamefont{J.~P.} \bibnamefont{Attfield}},
  \bibnamefont{and} \bibinfo{author}{\bibfnamefont{B.~D.} \bibnamefont{Gaulin}}
  (\bibinfo{year}{2014}), \eprint{cond-mat/1405.3172v1}.

\bibitem[{\citenamefont{Sato and Greedan}(1987)}]{Sato87}
\bibinfo{author}{\bibfnamefont{M.}~\bibnamefont{Sato}} \bibnamefont{and}
  \bibinfo{author}{\bibfnamefont{J.~E.} \bibnamefont{Greedan}},
  \bibinfo{journal}{Journal of Solid State Chemistry}
  \textbf{\bibinfo{volume}{67}}, \bibinfo{pages}{248 } (\bibinfo{year}{1987}),
  ISSN \bibinfo{issn}{0022-4596}.

\bibitem[{\citenamefont{Tam et~al.}(2010)\citenamefont{Tam, Hitchcock, and
  Gingras}}]{Tam10}
\bibinfo{author}{\bibfnamefont{K.-M.} \bibnamefont{Tam}},
  \bibinfo{author}{\bibfnamefont{A.~J.} \bibnamefont{Hitchcock}},
  \bibnamefont{and} \bibinfo{author}{\bibfnamefont{M.~J.~P.}
  \bibnamefont{Gingras}} (\bibinfo{year}{2010}), \eprint{cond-mat/1009.1272}.

\bibitem[{\citenamefont{Raju et~al.}(1992)\citenamefont{Raju, Gmelin, and
  Kremer}}]{Raju92}
\bibinfo{author}{\bibfnamefont{N.~P.} \bibnamefont{Raju}},
  \bibinfo{author}{\bibfnamefont{E.}~\bibnamefont{Gmelin}}, \bibnamefont{and}
  \bibinfo{author}{\bibfnamefont{R.~K.} \bibnamefont{Kremer}},
  \bibinfo{journal}{Phys. Rev. B} \textbf{\bibinfo{volume}{46}},
  \bibinfo{pages}{5405} (\bibinfo{year}{1992}).

\bibitem[{\citenamefont{Young and Katzgraber}(2004)}]{Young04}
\bibinfo{author}{\bibfnamefont{A.~P.} \bibnamefont{Young}} \bibnamefont{and}
  \bibinfo{author}{\bibfnamefont{H.~G.} \bibnamefont{Katzgraber}},
  \bibinfo{journal}{Phys. Rev. Lett.} \textbf{\bibinfo{volume}{93}},
  \bibinfo{pages}{207203} (\bibinfo{year}{2004}),
  \urlprefix\url{http://link.aps.org/doi/10.1103/PhysRevLett.93.207203}.

\bibitem[{\citenamefont{Sasaki et~al.}(2007)\citenamefont{Sasaki, Hukushima,
  Yoshino, and Takayama}}]{Sasaki07}
\bibinfo{author}{\bibfnamefont{M.}~\bibnamefont{Sasaki}},
  \bibinfo{author}{\bibfnamefont{K.}~\bibnamefont{Hukushima}},
  \bibinfo{author}{\bibfnamefont{H.}~\bibnamefont{Yoshino}}, \bibnamefont{and}
  \bibinfo{author}{\bibfnamefont{H.}~\bibnamefont{Takayama}},
  \bibinfo{journal}{Phys. Rev. Lett.} \textbf{\bibinfo{volume}{99}},
  \bibinfo{pages}{137202} (\bibinfo{year}{2007}),
  \urlprefix\url{http://link.aps.org/doi/10.1103/PhysRevLett.99.137202}.

\bibitem[{\citenamefont{Kino and Lüthi}(1971)}]{Kino71}
\bibinfo{author}{\bibfnamefont{Y.}~\bibnamefont{Kino}} \bibnamefont{and}
  \bibinfo{author}{\bibfnamefont{B.}~\bibnamefont{Lüthi}},
  \bibinfo{journal}{Solid State Communications} \textbf{\bibinfo{volume}{9}},
  \bibinfo{pages}{805 } (\bibinfo{year}{1971}), ISSN \bibinfo{issn}{0038-1098},
  \urlprefix\url{http://www.sciencedirect.com/science/article/pii/003810987190%
5680}.

\bibitem[{\citenamefont{Ratcliff et~al.}(2002)\citenamefont{Ratcliff, Lee,
  Broholm, Cheong, and Huang}}]{Ratcliff02}
\bibinfo{author}{\bibfnamefont{W.}~\bibnamefont{Ratcliff}},
  \bibinfo{author}{\bibfnamefont{S.-H.} \bibnamefont{Lee}},
  \bibinfo{author}{\bibfnamefont{C.}~\bibnamefont{Broholm}},
  \bibinfo{author}{\bibfnamefont{S.-W.} \bibnamefont{Cheong}},
  \bibnamefont{and} \bibinfo{author}{\bibfnamefont{Q.}~\bibnamefont{Huang}},
  \bibinfo{journal}{Phys. Rev. B} \textbf{\bibinfo{volume}{65}},
  \bibinfo{pages}{220406} (\bibinfo{year}{2002}).

\bibitem[{\citenamefont{Hanashima et~al.}(2013)\citenamefont{Hanashima, Kodama,
  Akahoshi, Kanadani, and Saito}}]{Hanashima13}
\bibinfo{author}{\bibfnamefont{K.}~\bibnamefont{Hanashima}},
  \bibinfo{author}{\bibfnamefont{Y.}~\bibnamefont{Kodama}},
  \bibinfo{author}{\bibfnamefont{D.}~\bibnamefont{Akahoshi}},
  \bibinfo{author}{\bibfnamefont{C.}~\bibnamefont{Kanadani}}, \bibnamefont{and}
  \bibinfo{author}{\bibfnamefont{T.}~\bibnamefont{Saito}},
  \bibinfo{journal}{Journal of the Physical Society of Japan}
  \textbf{\bibinfo{volume}{82}}, \bibinfo{pages}{024702}
  (\bibinfo{year}{2013}),
  \urlprefix\url{http://jpsj.ipap.jp/link?JPSJ/82/024702/}.

\bibitem[{\citenamefont{Booth et~al.}(2000)\citenamefont{Booth, Gardner, Kwei,
  Heffner, Bridges, and Subramanian}}]{Booth00}
\bibinfo{author}{\bibfnamefont{C.~H.} \bibnamefont{Booth}},
  \bibinfo{author}{\bibfnamefont{J.~S.} \bibnamefont{Gardner}},
  \bibinfo{author}{\bibfnamefont{G.~H.} \bibnamefont{Kwei}},
  \bibinfo{author}{\bibfnamefont{R.~H.} \bibnamefont{Heffner}},
  \bibinfo{author}{\bibfnamefont{F.}~\bibnamefont{Bridges}}, \bibnamefont{and}
  \bibinfo{author}{\bibfnamefont{M.~A.} \bibnamefont{Subramanian}},
  \bibinfo{journal}{Phys. Rev. B} \textbf{\bibinfo{volume}{62}},
  \bibinfo{pages}{R755} (\bibinfo{year}{2000}).

\bibitem[{\citenamefont{Greedan et~al.}(2009)\citenamefont{Greedan, Gout,
  Lozano-Gorrin, Derahkshan, Proffen, Kim, Bo\v{z}in, and
  Billinge}}]{Greedan09}
\bibinfo{author}{\bibfnamefont{J.~E.} \bibnamefont{Greedan}},
  \bibinfo{author}{\bibfnamefont{D.}~\bibnamefont{Gout}},
  \bibinfo{author}{\bibfnamefont{A.~D.} \bibnamefont{Lozano-Gorrin}},
  \bibinfo{author}{\bibfnamefont{S.}~\bibnamefont{Derahkshan}},
  \bibinfo{author}{\bibfnamefont{T.}~\bibnamefont{Proffen}},
  \bibinfo{author}{\bibfnamefont{H.-J.} \bibnamefont{Kim}},
  \bibinfo{author}{\bibfnamefont{E.}~\bibnamefont{Bo\v{z}in}},
  \bibnamefont{and} \bibinfo{author}{\bibfnamefont{S.~J.~L.}
  \bibnamefont{Billinge}}, \bibinfo{journal}{Phys. Rev. B}
  \textbf{\bibinfo{volume}{79}}, \bibinfo{pages}{014427}
  (\bibinfo{year}{2009}).

\bibitem[{\citenamefont{Keren and Gardner}(2001)}]{Keren01}
\bibinfo{author}{\bibfnamefont{A.}~\bibnamefont{Keren}} \bibnamefont{and}
  \bibinfo{author}{\bibfnamefont{J.~S.} \bibnamefont{Gardner}},
  \bibinfo{journal}{Phys. Rev. Lett.} \textbf{\bibinfo{volume}{87}},
  \bibinfo{pages}{177201} (\bibinfo{year}{2001}).

\bibitem[{\citenamefont{Ofer et~al.}(2010)\citenamefont{Ofer, Keren, Gardner,
  Ren, and MacFarlane}}]{Ofer10}
\bibinfo{author}{\bibfnamefont{O.}~\bibnamefont{Ofer}},
  \bibinfo{author}{\bibfnamefont{A.}~\bibnamefont{Keren}},
  \bibinfo{author}{\bibfnamefont{J.~S.} \bibnamefont{Gardner}},
  \bibinfo{author}{\bibfnamefont{Y.}~\bibnamefont{Ren}}, \bibnamefont{and}
  \bibinfo{author}{\bibfnamefont{W.~A.} \bibnamefont{MacFarlane}},
  \bibinfo{journal}{Phys. Rev. B} \textbf{\bibinfo{volume}{82}},
  \bibinfo{pages}{092403} (\bibinfo{year}{2010}),
  \urlprefix\url{http://link.aps.org/doi/10.1103/PhysRevB.82.092403}.

\bibitem[{\citenamefont{Sagi et~al.}(2005)\citenamefont{Sagi, Ofer, Keren, and
  Gardner}}]{Sagi05}
\bibinfo{author}{\bibfnamefont{E.}~\bibnamefont{Sagi}},
  \bibinfo{author}{\bibfnamefont{O.}~\bibnamefont{Ofer}},
  \bibinfo{author}{\bibfnamefont{A.}~\bibnamefont{Keren}}, \bibnamefont{and}
  \bibinfo{author}{\bibfnamefont{J.~S.} \bibnamefont{Gardner}},
  \bibinfo{journal}{Phys. Rev. Lett.} \textbf{\bibinfo{volume}{94}},
  \bibinfo{pages}{237202} (\bibinfo{year}{2005}).

\bibitem[{\citenamefont{Saunders and Chalker}(2008)}]{Saunders08}
\bibinfo{author}{\bibfnamefont{T.~E.} \bibnamefont{Saunders}} \bibnamefont{and}
  \bibinfo{author}{\bibfnamefont{J.~T.} \bibnamefont{Chalker}},
  \bibinfo{journal}{Phys. Rev. B} \textbf{\bibinfo{volume}{77}},
  \bibinfo{pages}{214438} (\bibinfo{year}{2008}).

\bibitem[{\citenamefont{Shinaoka
  et~al.}(2011{\natexlab{a}})\citenamefont{Shinaoka, Tomita, and
  Motome}}]{Shinaoka10b}
\bibinfo{author}{\bibfnamefont{H.}~\bibnamefont{Shinaoka}},
  \bibinfo{author}{\bibfnamefont{Y.}~\bibnamefont{Tomita}}, \bibnamefont{and}
  \bibinfo{author}{\bibfnamefont{Y.}~\bibnamefont{Motome}},
  \bibinfo{journal}{Phys. Rev. Lett.} \textbf{\bibinfo{volume}{107}},
  \bibinfo{pages}{047204} (\bibinfo{year}{2011}{\natexlab{a}}),
  \urlprefix\url{http://link.aps.org/doi/10.1103/PhysRevLett.107.047204}.

\bibitem[{\citenamefont{Shinaoka et~al.}(2012)\citenamefont{Shinaoka, Tomita,
  and Motome}}]{Shinaoka-LT2011}
\bibinfo{author}{\bibfnamefont{H.}~\bibnamefont{Shinaoka}},
  \bibinfo{author}{\bibfnamefont{Y.}~\bibnamefont{Tomita}}, \bibnamefont{and}
  \bibinfo{author}{\bibfnamefont{Y.}~\bibnamefont{Motome}},
  \bibinfo{journal}{Journal of Physics: Conference Series}
  \textbf{\bibinfo{volume}{400}}, \bibinfo{pages}{032087}
  (\bibinfo{year}{2012}),
  \urlprefix\url{http://stacks.iop.org/1742-6596/400/i=3/a=032087}.

\bibitem[{\citenamefont{Tchernyshyov
  et~al.}(2002{\natexlab{a}})\citenamefont{Tchernyshyov, Moessner, and
  Sondhi}}]{Tchernyshyov02}
\bibinfo{author}{\bibfnamefont{O.}~\bibnamefont{Tchernyshyov}},
  \bibinfo{author}{\bibfnamefont{R.}~\bibnamefont{Moessner}}, \bibnamefont{and}
  \bibinfo{author}{\bibfnamefont{S.~L.} \bibnamefont{Sondhi}},
  \bibinfo{journal}{Phys. Rev. Lett.} \textbf{\bibinfo{volume}{88}},
  \bibinfo{pages}{067203} (\bibinfo{year}{2002}{\natexlab{a}}),
  \urlprefix\url{http://link.aps.org/doi/10.1103/PhysRevLett.88.067203}.

\bibitem[{\citenamefont{Tchernyshyov
  et~al.}(2002{\natexlab{b}})\citenamefont{Tchernyshyov, Moessner, and
  Sondhi}}]{Tchernyshyov02b}
\bibinfo{author}{\bibfnamefont{O.}~\bibnamefont{Tchernyshyov}},
  \bibinfo{author}{\bibfnamefont{R.}~\bibnamefont{Moessner}}, \bibnamefont{and}
  \bibinfo{author}{\bibfnamefont{S.~L.} \bibnamefont{Sondhi}},
  \bibinfo{journal}{Phys. Rev. B} \textbf{\bibinfo{volume}{66}},
  \bibinfo{pages}{064403} (\bibinfo{year}{2002}{\natexlab{b}}),
  \urlprefix\url{http://link.aps.org/doi/10.1103/PhysRevB.66.064403}.

\bibitem[{\citenamefont{Bergman et~al.}(2006)\citenamefont{Bergman, Shindou,
  Fiete, and Balents}}]{Bergman06}
\bibinfo{author}{\bibfnamefont{D.~L.} \bibnamefont{Bergman}},
  \bibinfo{author}{\bibfnamefont{R.}~\bibnamefont{Shindou}},
  \bibinfo{author}{\bibfnamefont{G.~A.} \bibnamefont{Fiete}}, \bibnamefont{and}
  \bibinfo{author}{\bibfnamefont{L.}~\bibnamefont{Balents}},
  \bibinfo{journal}{Phys. Rev. B} \textbf{\bibinfo{volume}{74}},
  \bibinfo{pages}{134409} (\bibinfo{year}{2006}),
  \urlprefix\url{http://link.aps.org/doi/10.1103/PhysRevB.74.134409}.

\bibitem[{\citenamefont{Ji et~al.}(2009)\citenamefont{Ji, Lee, Broholm, Koo,
  Ratcliff, Cheong, and Zschack}}]{Ji09}
\bibinfo{author}{\bibfnamefont{S.}~\bibnamefont{Ji}},
  \bibinfo{author}{\bibfnamefont{S.-H.} \bibnamefont{Lee}},
  \bibinfo{author}{\bibfnamefont{C.}~\bibnamefont{Broholm}},
  \bibinfo{author}{\bibfnamefont{T.~Y.} \bibnamefont{Koo}},
  \bibinfo{author}{\bibfnamefont{W.}~\bibnamefont{Ratcliff}},
  \bibinfo{author}{\bibfnamefont{S.-W.} \bibnamefont{Cheong}},
  \bibnamefont{and} \bibinfo{author}{\bibfnamefont{P.}~\bibnamefont{Zschack}},
  \bibinfo{journal}{Phys. Rev. Lett.} \textbf{\bibinfo{volume}{103}},
  \bibinfo{pages}{037201} (\bibinfo{year}{2009}),
  \urlprefix\url{http://link.aps.org/doi/10.1103/PhysRevLett.103.037201}.

\bibitem[{\citenamefont{Ueda et~al.}(2006)\citenamefont{Ueda, Mitamura, Goto,
  and Ueda}}]{Ueda06}
\bibinfo{author}{\bibfnamefont{H.}~\bibnamefont{Ueda}},
  \bibinfo{author}{\bibfnamefont{H.}~\bibnamefont{Mitamura}},
  \bibinfo{author}{\bibfnamefont{T.}~\bibnamefont{Goto}}, \bibnamefont{and}
  \bibinfo{author}{\bibfnamefont{Y.}~\bibnamefont{Ueda}},
  \bibinfo{journal}{Phys. Rev. B} \textbf{\bibinfo{volume}{73}},
  \bibinfo{pages}{094415} (\bibinfo{year}{2006}),
  \urlprefix\url{http://link.aps.org/doi/10.1103/PhysRevB.73.094415}.

\bibitem[{\citenamefont{Matsuda et~al.}(2007)\citenamefont{Matsuda, Ueda,
  Kikkawa, Tanaka, Katsumata, Narumi, Inami, and Ueda}}]{Matsuda07}
\bibinfo{author}{\bibfnamefont{M.}~\bibnamefont{Matsuda}},
  \bibinfo{author}{\bibfnamefont{H.}~\bibnamefont{Ueda}},
  \bibinfo{author}{\bibfnamefont{A.}~\bibnamefont{Kikkawa}},
  \bibinfo{author}{\bibfnamefont{Y.}~\bibnamefont{Tanaka}},
  \bibinfo{author}{\bibfnamefont{K.}~\bibnamefont{Katsumata}},
  \bibinfo{author}{\bibfnamefont{Y.}~\bibnamefont{Narumi}},
  \bibinfo{author}{\bibfnamefont{T.}~\bibnamefont{Inami}}, \bibnamefont{and}
  \bibinfo{author}{\bibfnamefont{Y.}~\bibnamefont{Ueda}},
  \bibinfo{journal}{Nature Phys.} \textbf{\bibinfo{volume}{3}},
  \bibinfo{pages}{397} (\bibinfo{year}{2007}).

\bibitem[{\citenamefont{Matsuda}(2007)}]{Matsuda20077}
\bibinfo{author}{\bibfnamefont{M.}~\bibnamefont{Matsuda}},
  \bibinfo{journal}{Physica B: Condensed Matter}
  \textbf{\bibinfo{volume}{397}}, \bibinfo{pages}{7 } (\bibinfo{year}{2007}),
  ISSN \bibinfo{issn}{0921-4526}.

\bibitem[{\citenamefont{Moessner and
  Chalker}(1998{\natexlab{b}})}]{Moessner98b}
\bibinfo{author}{\bibfnamefont{R.}~\bibnamefont{Moessner}} \bibnamefont{and}
  \bibinfo{author}{\bibfnamefont{J.~T.} \bibnamefont{Chalker}},
  \bibinfo{journal}{Phys. Rev. B} \textbf{\bibinfo{volume}{58}},
  \bibinfo{pages}{12049} (\bibinfo{year}{1998}{\natexlab{b}}).

\bibitem[{\citenamefont{Shannon et~al.}(2010)\citenamefont{Shannon, Penc, and
  Motome}}]{Shannon10}
\bibinfo{author}{\bibfnamefont{N.}~\bibnamefont{Shannon}},
  \bibinfo{author}{\bibfnamefont{K.}~\bibnamefont{Penc}}, \bibnamefont{and}
  \bibinfo{author}{\bibfnamefont{Y.}~\bibnamefont{Motome}},
  \bibinfo{journal}{Phys. Rev. B} \textbf{\bibinfo{volume}{81}},
  \bibinfo{pages}{184409} (\bibinfo{year}{2010}).

\bibitem[{\citenamefont{Bernal and Fowlers}(1933)}]{Bernal33}
\bibinfo{author}{\bibfnamefont{J.~D.} \bibnamefont{Bernal}} \bibnamefont{and}
  \bibinfo{author}{\bibfnamefont{R.~H.} \bibnamefont{Fowlers}},
  \bibinfo{journal}{J. Chem. Phys.} \textbf{\bibinfo{volume}{1}},
  \bibinfo{pages}{515} (\bibinfo{year}{1933}).

\bibitem[{\citenamefont{Pauling}(1935)}]{Pauling35}
\bibinfo{author}{\bibfnamefont{L.}~\bibnamefont{Pauling}}, \bibinfo{journal}{J.
  Am. Chem. Soc.} \textbf{\bibinfo{volume}{57}}, \bibinfo{pages}{2680}
  (\bibinfo{year}{1935}).

\bibitem[{\citenamefont{Reimers et~al.}(1991)\citenamefont{Reimers, Berlinsky,
  and Shi}}]{Reimers91}
\bibinfo{author}{\bibfnamefont{J.~N.} \bibnamefont{Reimers}},
  \bibinfo{author}{\bibfnamefont{A.~J.} \bibnamefont{Berlinsky}},
  \bibnamefont{and} \bibinfo{author}{\bibfnamefont{A.-C.} \bibnamefont{Shi}},
  \bibinfo{journal}{Phys. Rev. B} \textbf{\bibinfo{volume}{43}},
  \bibinfo{pages}{865} (\bibinfo{year}{1991}),
  \urlprefix\url{http://link.aps.org/doi/10.1103/PhysRevB.43.865}.

\bibitem[{\citenamefont{Chern et~al.}(2008)\citenamefont{Chern, Moessner, and
  Tchernyshyov}}]{Chern08}
\bibinfo{author}{\bibfnamefont{G.-W.} \bibnamefont{Chern}},
  \bibinfo{author}{\bibfnamefont{R.}~\bibnamefont{Moessner}}, \bibnamefont{and}
  \bibinfo{author}{\bibfnamefont{O.}~\bibnamefont{Tchernyshyov}},
  \bibinfo{journal}{Phys. Rev. B} \textbf{\bibinfo{volume}{78}},
  \bibinfo{pages}{144418} (\bibinfo{year}{2008}).

\bibitem[{\citenamefont{Marsaglia}(1972)}]{Marsaglia72}
\bibinfo{author}{\bibfnamefont{G.}~\bibnamefont{Marsaglia}},
  \bibinfo{journal}{The Annals of Mathematical Statistics}
  \textbf{\bibinfo{volume}{43}}, \bibinfo{pages}{645} (\bibinfo{year}{1972}).

\bibitem[{\citenamefont{Alonso et~al.}(1996)\citenamefont{Alonso, Taranc\'on,
  Ballesteros, Fern\'andez, Mart\'\i{}n-Mayor, and Mu\~noz Sudupe}}]{Alonso96}
\bibinfo{author}{\bibfnamefont{J.~L.} \bibnamefont{Alonso}},
  \bibinfo{author}{\bibfnamefont{A.}~\bibnamefont{Taranc\'on}},
  \bibinfo{author}{\bibfnamefont{H.~G.} \bibnamefont{Ballesteros}},
  \bibinfo{author}{\bibfnamefont{L.~A.} \bibnamefont{Fern\'andez}},
  \bibinfo{author}{\bibfnamefont{V.}~\bibnamefont{Mart\'\i{}n-Mayor}},
  \bibnamefont{and} \bibinfo{author}{\bibfnamefont{A.}~\bibnamefont{Mu\~noz
  Sudupe}}, \bibinfo{journal}{Phys. Rev. B} \textbf{\bibinfo{volume}{53}},
  \bibinfo{pages}{2537} (\bibinfo{year}{1996}).

\bibitem[{\citenamefont{Hukushima and Nemoto}(1996)}]{Hukushima96}
\bibinfo{author}{\bibfnamefont{K.}~\bibnamefont{Hukushima}} \bibnamefont{and}
  \bibinfo{author}{\bibfnamefont{K.}~\bibnamefont{Nemoto}},
  \bibinfo{journal}{Journal of the Physical Society of Japan}
  \textbf{\bibinfo{volume}{65}}, \bibinfo{pages}{1604} (\bibinfo{year}{1996}).

\bibitem[{\citenamefont{Shinaoka and Motome}(2010)}]{Shinaoka-LM1}
\bibinfo{author}{\bibfnamefont{H.}~\bibnamefont{Shinaoka}} \bibnamefont{and}
  \bibinfo{author}{\bibfnamefont{Y.}~\bibnamefont{Motome}},
  \bibinfo{journal}{Phys. Rev. B} \textbf{\bibinfo{volume}{82}},
  \bibinfo{pages}{134420} (\bibinfo{year}{2010}).

\bibitem[{\citenamefont{Shinaoka
  et~al.}(2011{\natexlab{b}})\citenamefont{Shinaoka, Motome, and
  Tomita}}]{Shinaoka-LM2}
\bibinfo{author}{\bibfnamefont{H.}~\bibnamefont{Shinaoka}},
  \bibinfo{author}{\bibfnamefont{Y.}~\bibnamefont{Motome}}, \bibnamefont{and}
  \bibinfo{author}{\bibfnamefont{Y.}~\bibnamefont{Tomita}},
  \bibinfo{journal}{J. Phys.: Conf. Ser.} \textbf{\bibinfo{volume}{320}},
  \bibinfo{pages}{012009} (\bibinfo{year}{2011}{\natexlab{b}}).

\bibitem[{\citenamefont{Melko and Gingras}(2004)}]{Melko04}
\bibinfo{author}{\bibfnamefont{R.~G.} \bibnamefont{Melko}} \bibnamefont{and}
  \bibinfo{author}{\bibfnamefont{M.~J.~P.} \bibnamefont{Gingras}},
  \bibinfo{journal}{Journal of Physics: Condensed Matter}
  \textbf{\bibinfo{volume}{16}}, \bibinfo{pages}{R1277} (\bibinfo{year}{2004}),
  \urlprefix\url{http://stacks.iop.org/0953-8984/16/i=43/a=R02}.

\bibitem[{\citenamefont{Melko et~al.}(2001)\citenamefont{Melko, den Hertog, and
  Gingras}}]{Melko01}
\bibinfo{author}{\bibfnamefont{R.~G.} \bibnamefont{Melko}},
  \bibinfo{author}{\bibfnamefont{B.~C.} \bibnamefont{den Hertog}},
  \bibnamefont{and} \bibinfo{author}{\bibfnamefont{M.~J.~P.}
  \bibnamefont{Gingras}}, \bibinfo{journal}{Phys. Rev. Lett.}
  \textbf{\bibinfo{volume}{87}}, \bibinfo{pages}{067203}
  (\bibinfo{year}{2001}).

\bibitem[{\citenamefont{Wang et~al.}(2012)\citenamefont{Wang, De~Sterck, and
  Melko}}]{Wang12}
\bibinfo{author}{\bibfnamefont{Y.}~\bibnamefont{Wang}},
  \bibinfo{author}{\bibfnamefont{H.}~\bibnamefont{De~Sterck}},
  \bibnamefont{and} \bibinfo{author}{\bibfnamefont{R.~G.} \bibnamefont{Melko}},
  \bibinfo{journal}{Phys. Rev. E} \textbf{\bibinfo{volume}{85}},
  \bibinfo{pages}{036704} (\bibinfo{year}{2012}),
  \urlprefix\url{http://link.aps.org/doi/10.1103/PhysRevE.85.036704}.

\bibitem[{\citenamefont{Anderson}(1956)}]{Anderson56}
\bibinfo{author}{\bibfnamefont{P.~W.} \bibnamefont{Anderson}},
  \bibinfo{journal}{Phys. Rev.} \textbf{\bibinfo{volume}{102}},
  \bibinfo{pages}{1008} (\bibinfo{year}{1956}).

\bibitem[{\citenamefont{Isakov et~al.}(2004)\citenamefont{Isakov, Raman,
  Moessner, and Sondhi}}]{Isakov04}
\bibinfo{author}{\bibfnamefont{S.~V.} \bibnamefont{Isakov}},
  \bibinfo{author}{\bibfnamefont{K.~S.} \bibnamefont{Raman}},
  \bibinfo{author}{\bibfnamefont{R.}~\bibnamefont{Moessner}}, \bibnamefont{and}
  \bibinfo{author}{\bibfnamefont{S.~L.} \bibnamefont{Sondhi}},
  \bibinfo{journal}{Phys. Rev. B} \textbf{\bibinfo{volume}{70}},
  \bibinfo{pages}{104418} (\bibinfo{year}{2004}).

\bibitem[{\citenamefont{Ruff et~al.}(2005)\citenamefont{Ruff, Melko, and
  Gingras}}]{Jacob05}
\bibinfo{author}{\bibfnamefont{J.~P.~C.} \bibnamefont{Ruff}},
  \bibinfo{author}{\bibfnamefont{R.~G.} \bibnamefont{Melko}}, \bibnamefont{and}
  \bibinfo{author}{\bibfnamefont{M.~J.~P.} \bibnamefont{Gingras}},
  \bibinfo{journal}{Phys. Rev. Lett.} \textbf{\bibinfo{volume}{95}},
  \bibinfo{pages}{097202} (\bibinfo{year}{2005}).

\bibitem[{\citenamefont{{Jaubert} et~al.}(2010)\citenamefont{{Jaubert},
  {Chalker}, {Holdsworth}, and {Moessner}}}]{Jaubert10}
\bibinfo{author}{\bibfnamefont{L.~D.~C.} \bibnamefont{{Jaubert}}},
  \bibinfo{author}{\bibfnamefont{J.~T.} \bibnamefont{{Chalker}}},
  \bibinfo{author}{\bibfnamefont{P.~C.~W.} \bibnamefont{{Holdsworth}}},
  \bibnamefont{and}
  \bibinfo{author}{\bibfnamefont{R.}~\bibnamefont{{Moessner}}}
  (\bibinfo{year}{2010}), \eprint{cond-mat/1003.4896v1}.

\bibitem[{\citenamefont{Creutz}(1987)}]{Michael87}
\bibinfo{author}{\bibfnamefont{M.}~\bibnamefont{Creutz}},
  \bibinfo{journal}{Phys. Rev. D} \textbf{\bibinfo{volume}{36}},
  \bibinfo{pages}{515} (\bibinfo{year}{1987}).

\bibitem[{\citenamefont{Landau and Binder}(2009)}]{LandauMC}
\bibinfo{author}{\bibfnamefont{D.~P.} \bibnamefont{Landau}} \bibnamefont{and}
  \bibinfo{author}{\bibfnamefont{K.}~\bibnamefont{Binder}},
  \emph{\bibinfo{title}{A Guide to Monte Carlo Simulations in Statistical
  Physics (Third Edition)}} (\bibinfo{publisher}{Cambridge University Press},
  \bibinfo{year}{2009}).

\bibitem[{\citenamefont{Edwards and Anderson}(1975)}]{Edwards75}
\bibinfo{author}{\bibfnamefont{S.~F.} \bibnamefont{Edwards}} \bibnamefont{and}
  \bibinfo{author}{\bibfnamefont{P.~W.} \bibnamefont{Anderson}},
  \bibinfo{journal}{J. Phys. F: Met. Phys.} \textbf{\bibinfo{volume}{5}},
  \bibinfo{pages}{965} (\bibinfo{year}{1975}).

\bibitem[{\citenamefont{K\"{o}bler et~al.}(1996)\citenamefont{K\"{o}bler,
  Mueller, Smardz, Maier, Fischer, Olefs, and Zinn}}]{Kobler96}
\bibinfo{author}{\bibfnamefont{U.}~\bibnamefont{K\"{o}bler}},
  \bibinfo{author}{\bibfnamefont{R.}~\bibnamefont{Mueller}},
  \bibinfo{author}{\bibfnamefont{L.}~\bibnamefont{Smardz}},
  \bibinfo{author}{\bibfnamefont{D.}~\bibnamefont{Maier}},
  \bibinfo{author}{\bibfnamefont{K.}~\bibnamefont{Fischer}},
  \bibinfo{author}{\bibfnamefont{B.}~\bibnamefont{Olefs}}, \bibnamefont{and}
  \bibinfo{author}{\bibfnamefont{W.}~\bibnamefont{Zinn}}, \bibinfo{journal}{Z.
  Phys. B} \textbf{\bibinfo{volume}{100}}, \bibinfo{pages}{497}
  (\bibinfo{year}{1996}).

\bibitem[{\citenamefont{Nakamura and ichi Endoh}(2002)}]{Nakamura02}
\bibinfo{author}{\bibfnamefont{T.}~\bibnamefont{Nakamura}} \bibnamefont{and}
  \bibinfo{author}{\bibfnamefont{S.}~\bibnamefont{ichi Endoh}},
  \bibinfo{journal}{Journal of the Physical Society of Japan}
  \textbf{\bibinfo{volume}{71}}, \bibinfo{pages}{2113} (\bibinfo{year}{2002}),
  \urlprefix\url{http://jpsj.ipap.jp/link?JPSJ/71/2113/}.

\bibitem[{\citenamefont{Fernandez et~al.}(2009)\citenamefont{Fernandez,
  Martin-Mayor, Perez-Gaviro, Tarancon, and Young}}]{Fernandez09}
\bibinfo{author}{\bibfnamefont{L.~A.} \bibnamefont{Fernandez}},
  \bibinfo{author}{\bibfnamefont{V.}~\bibnamefont{Martin-Mayor}},
  \bibinfo{author}{\bibfnamefont{S.}~\bibnamefont{Perez-Gaviro}},
  \bibinfo{author}{\bibfnamefont{A.}~\bibnamefont{Tarancon}}, \bibnamefont{and}
  \bibinfo{author}{\bibfnamefont{A.~P.} \bibnamefont{Young}},
  \bibinfo{journal}{Phys. Rev. B} \textbf{\bibinfo{volume}{80}},
  \bibinfo{pages}{024422} (\bibinfo{year}{2009}).

\bibitem[{\citenamefont{Ramirez et~al.}(1992)\citenamefont{Ramirez, Coleman,
  Chandra, Br\"uck, Menovsky, Fisk, and Bucher}}]{Ramirez92}
\bibinfo{author}{\bibfnamefont{A.~P.} \bibnamefont{Ramirez}},
  \bibinfo{author}{\bibfnamefont{P.}~\bibnamefont{Coleman}},
  \bibinfo{author}{\bibfnamefont{P.}~\bibnamefont{Chandra}},
  \bibinfo{author}{\bibfnamefont{E.}~\bibnamefont{Br\"uck}},
  \bibinfo{author}{\bibfnamefont{A.~A.} \bibnamefont{Menovsky}},
  \bibinfo{author}{\bibfnamefont{Z.}~\bibnamefont{Fisk}}, \bibnamefont{and}
  \bibinfo{author}{\bibfnamefont{E.}~\bibnamefont{Bucher}},
  \bibinfo{journal}{Phys. Rev. Lett.} \textbf{\bibinfo{volume}{68}},
  \bibinfo{pages}{2680} (\bibinfo{year}{1992}),
  \urlprefix\url{http://link.aps.org/doi/10.1103/PhysRevLett.68.2680}.

\bibitem[{\citenamefont{Fisher and Huse}(1988)}]{Fisher88}
\bibinfo{author}{\bibfnamefont{D.~S.} \bibnamefont{Fisher}} \bibnamefont{and}
  \bibinfo{author}{\bibfnamefont{D.~A.} \bibnamefont{Huse}},
  \bibinfo{journal}{Phys. Rev. B} \textbf{\bibinfo{volume}{38}},
  \bibinfo{pages}{386} (\bibinfo{year}{1988}),
  \urlprefix\url{http://link.aps.org/doi/10.1103/PhysRevB.38.386}.

\bibitem[{\citenamefont{Chalupa}(1977)}]{Chalupa77}
\bibinfo{author}{\bibfnamefont{J.}~\bibnamefont{Chalupa}},
  \bibinfo{journal}{Solid State Communications} \textbf{\bibinfo{volume}{22}},
  \bibinfo{pages}{315 } (\bibinfo{year}{1977}), ISSN \bibinfo{issn}{0038-1098},
  \urlprefix\url{http://www.sciencedirect.com/science/article/pii/003810987791%
4399}.

\bibitem[{\citenamefont{den Hertog and Gingras}(2000)}]{Hertog00}
\bibinfo{author}{\bibfnamefont{B.~C.} \bibnamefont{den Hertog}}
  \bibnamefont{and} \bibinfo{author}{\bibfnamefont{M.~J.~P.}
  \bibnamefont{Gingras}}, \bibinfo{journal}{Phys. Rev. Lett.}
  \textbf{\bibinfo{volume}{84}}, \bibinfo{pages}{3430} (\bibinfo{year}{2000}).

\bibitem[{\citenamefont{Harris et~al.}(1997)\citenamefont{Harris, Bramwell,
  McMorrow, Zeiske, and Godfrey}}]{Harris97}
\bibinfo{author}{\bibfnamefont{M.~J.} \bibnamefont{Harris}},
  \bibinfo{author}{\bibfnamefont{S.~T.} \bibnamefont{Bramwell}},
  \bibinfo{author}{\bibfnamefont{D.~F.} \bibnamefont{McMorrow}},
  \bibinfo{author}{\bibfnamefont{T.}~\bibnamefont{Zeiske}}, \bibnamefont{and}
  \bibinfo{author}{\bibfnamefont{K.~W.} \bibnamefont{Godfrey}},
  \bibinfo{journal}{Phys. Rev. Lett.} \textbf{\bibinfo{volume}{79}},
  \bibinfo{pages}{2554} (\bibinfo{year}{1997}).

\bibitem[{\citenamefont{Ramirez et~al.}(1999)\citenamefont{Ramirez, Hayashi,
  Cava, Siddharthan, and Shastry}}]{Ramirez99}
\bibinfo{author}{\bibfnamefont{A.~P.} \bibnamefont{Ramirez}},
  \bibinfo{author}{\bibfnamefont{A.}~\bibnamefont{Hayashi}},
  \bibinfo{author}{\bibfnamefont{R.~J.} \bibnamefont{Cava}},
  \bibinfo{author}{\bibfnamefont{R.}~\bibnamefont{Siddharthan}},
  \bibnamefont{and} \bibinfo{author}{\bibfnamefont{B.~S.}
  \bibnamefont{Shastry}}, \bibinfo{journal}{Nature}
  \textbf{\bibinfo{volume}{399}}, \bibinfo{pages}{333} (\bibinfo{year}{1999}).

\bibitem[{\citenamefont{Bramwell and Gingras}(2001)}]{Bramwell01}
\bibinfo{author}{\bibfnamefont{S.~T.} \bibnamefont{Bramwell}} \bibnamefont{and}
  \bibinfo{author}{\bibfnamefont{M.~J.~P.} \bibnamefont{Gingras}},
  \bibinfo{journal}{Science} \textbf{\bibinfo{volume}{294}},
  \bibinfo{pages}{1495} (\bibinfo{year}{2001}),
  \eprint{http://www.sciencemag.org/cgi/reprint/294/5546/1495.pdf},
  \urlprefix\url{http://www.sciencemag.org/cgi/content/abstract/294/5546/1495}.

\bibitem[{\citenamefont{Castelnovo et~al.}(2011)\citenamefont{Castelnovo,
  Moessner, and Sondhi}}]{Castelnovo11}
\bibinfo{author}{\bibfnamefont{C.}~\bibnamefont{Castelnovo}},
  \bibinfo{author}{\bibfnamefont{R.}~\bibnamefont{Moessner}}, \bibnamefont{and}
  \bibinfo{author}{\bibfnamefont{S.~L.} \bibnamefont{Sondhi}},
  \bibinfo{journal}{Phys. Rev. B} \textbf{\bibinfo{volume}{84}},
  \bibinfo{pages}{144435} (\bibinfo{year}{2011}),
  \urlprefix\url{http://link.aps.org/doi/10.1103/PhysRevB.84.144435}.

\bibitem[{\citenamefont{Jaubert and Holdsworth}(2009)}]{Jaubert09}
\bibinfo{author}{\bibfnamefont{L.~D.~C.} \bibnamefont{Jaubert}}
  \bibnamefont{and} \bibinfo{author}{\bibfnamefont{P.~C.~W.}
  \bibnamefont{Holdsworth}}, \bibinfo{journal}{Nature Phys.}
  \textbf{\bibinfo{volume}{5}}, \bibinfo{pages}{258} (\bibinfo{year}{2009}).

\bibitem[{\citenamefont{Castelnovo et~al.}(2010)\citenamefont{Castelnovo,
  Moessner, and Sondhi}}]{Castelnovo10}
\bibinfo{author}{\bibfnamefont{C.}~\bibnamefont{Castelnovo}},
  \bibinfo{author}{\bibfnamefont{R.}~\bibnamefont{Moessner}}, \bibnamefont{and}
  \bibinfo{author}{\bibfnamefont{S.~L.} \bibnamefont{Sondhi}},
  \bibinfo{journal}{Phys. Rev. Lett.} \textbf{\bibinfo{volume}{104}},
  \bibinfo{pages}{107201} (\bibinfo{year}{2010}),
  \urlprefix\url{http://link.aps.org/doi/10.1103/PhysRevLett.104.107201}.

\bibitem[{\citenamefont{Roy et~al.}()\citenamefont{Roy, Pandey, Zhang,
  Heitmann, Vaknin, Johnston, and Furukawa}}]{Roy13}
\bibinfo{author}{\bibfnamefont{B.}~\bibnamefont{Roy}},
  \bibinfo{author}{\bibfnamefont{A.}~\bibnamefont{Pandey}},
  \bibinfo{author}{\bibfnamefont{Q.}~\bibnamefont{Zhang}},
  \bibinfo{author}{\bibfnamefont{T.~W.} \bibnamefont{Heitmann}},
  \bibinfo{author}{\bibfnamefont{D.}~\bibnamefont{Vaknin}},
  \bibinfo{author}{\bibfnamefont{D.~C.} \bibnamefont{Johnston}},
  \bibnamefont{and} \bibinfo{author}{\bibfnamefont{Y.}~\bibnamefont{Furukawa}}.

\bibitem[{\citenamefont{Shinaoka et~al.}(2013)\citenamefont{Shinaoka, Motome,
  Miyake, and Ishibashi}}]{Shinaoka-YMO}
\bibinfo{author}{\bibfnamefont{H.}~\bibnamefont{Shinaoka}},
  \bibinfo{author}{\bibfnamefont{Y.}~\bibnamefont{Motome}},
  \bibinfo{author}{\bibfnamefont{T.}~\bibnamefont{Miyake}}, \bibnamefont{and}
  \bibinfo{author}{\bibfnamefont{S.}~\bibnamefont{Ishibashi}},
  \bibinfo{journal}{Phys. Rev. B} \textbf{\bibinfo{volume}{88}},
  \bibinfo{pages}{174422} (\bibinfo{year}{2013}),
  \urlprefix\url{http://link.aps.org/doi/10.1103/PhysRevB.88.174422}.

\end{thebibliography}
